\documentclass[lettersize,journal]{IEEEtran}
\usepackage{lineno}
\usepackage{cite}
\usepackage{hyperref}
\usepackage{amsmath,amssymb,amsfonts}
\usepackage{amsmath}
\usepackage{amsthm}
\DeclareMathOperator*{\argmax}{argmax}
\usepackage{algpseudocode}
\usepackage{graphicx}
\usepackage{textcomp}
\usepackage{xcolor}
\usepackage{graphicx}
\usepackage{float}
\usepackage{stfloats}
\usepackage{subfigure}
\usepackage{amsmath}
\usepackage{amsfonts,amssymb}
\usepackage{mathrsfs}
\usepackage{mathtools}
\usepackage{algorithm}
\usepackage{algorithmicx}
\usepackage{algpseudocode}
\usepackage{bm}
\usepackage{multirow}
\usepackage{array}
\usepackage{amssymb}
\usepackage{amsmath}
\usepackage{cite}
\usepackage{url}
\usepackage{xcolor}
\usepackage{cite,graphicx,amsmath,amssymb}
\usepackage{subfigure}
\usepackage{fancyhdr}
\usepackage{mdwmath}
\usepackage{mdwtab}
\usepackage{caption}
\usepackage{amsthm}
\usepackage{setspace}
\usepackage{bm}
\usepackage{algorithm}
\usepackage{algpseudocode}
\usepackage{mathtools}
\usepackage{dsfont}
\usepackage{bbm}
\usepackage{cases}
\usepackage{booktabs}
\newtheorem{remark}{Remark}
\theoremstyle{definition}
\newtheorem{theorem}{Theorem}

\newtheorem{lemma}{Lemma}

\newtheorem{corollary}{Corollary}

\makeatletter
\newcommand{\biggg}{\bBigg@{3}}
\newcommand{\Biggg}{\bBigg@{3.5}}
\makeatother
\begin{document}

\title{Physical Layer Security for Pinching-Antenna Systems (PASS)}
\author{
\author{
Mingjun Sun,~\IEEEmembership{Graduate Student Member,~IEEE,} Chongjun Ouyang,~\IEEEmembership{Member,~IEEE,}\\ Shaochuan Wu,~\IEEEmembership{Senior Member,~IEEE,} and Yuanwei Liu,~\IEEEmembership{Fellow,~IEEE}
\thanks{Mingjun Sun and Shaochuan Wu are with the School of Electronics and Information Engineering, Harbin Institute of Technology, Harbin 150001, China (e-mail: sunmj@stu.hit.edu.cn; scwu@hit.edu.cn).}
\thanks{Chongjun Ouyang is with the School of Electronic Engineering and Computer Science, Queen Mary University of London, London, E1 4NS, U.K. (e-mail: c.ouyang@qmul.ac.uk).}
\thanks{Yuanwei Liu is with the Department of Electrical and Electronic Engineering, The University of Hong Kong, Hong Kong (e-mail: yuanwei@hku.hk).}
}}

%

\maketitle

\begin{abstract}
The pinching-antenna system (PASS) introduces new degrees of freedom (DoFs) for physical layer security (PLS) through \textit{pinching beamforming}.
In this paper, a couple of scenarios for secure beamforming for PASS are studied. 1) For the case with a single legitimate user (Bob) and a single eavesdropper (Eve), a closed-form expression for the optimal baseband beamformer is derived. On this basis, a gradient-based method is proposed to optimize the activated positions of pinching antennas (PAs). 2) For the case with multiple Bobs and multiple Eves, a fractional programming (FP)-based block coordinate descent (BCD) algorithm, termed FP-BCD, is proposed for optimizing the weighted secrecy sum-rate (WSSR). 
Specifically, a closed-form baseband beamformer is obtained via Lagrange multiplier method. Furthermore, owing to the non-convex objective function exhibiting numerous stationary points, a low-complexity one-dimensional search is used to find a high-quality solution of the PAs' activated locations. Numerical results are provided to demonstrate that: i) All proposed algorithms achieve stable convergence within a few iterations, 
ii) across all considered power ranges, the FP-BCD algorithm outperforms baseline methods using zero-forcing (ZF) and maximal-ratio transmission (MRT) beamforming in terms of the WSSR,
and iii) PASS achieves a significantly higher secrecy rate than traditional fixed-antenna systems.
\end{abstract}


\section{Introduction}
\IEEEPARstart{M}{ultiple-antenna} technology is one of the key techniques for improving the spectral efficiency of wireless communication systems. By increasing spatial degrees of freedom (DoFs), it enables transmit beamforming to enhance received signal strength while effectively suppressing interference, thereby improving channel capacity \cite{ref101, ref30}. A successful example of this is the massive multiple-input multiple-output (Massive MIMO) technology, which has been incorporated into the 5G standard and was commercially deployed in 2018\cite{ref2}. 
Recently, multiple-antenna technology has seen significant advancements, offering new possibilities for providing enhanced communication efficiency. To customize wireless channels, various flexible-antenna systems have been proposed, such as reconfigurable intelligent surfaces (RISs)\cite{ref108}, movable antennas\cite{ref109}, and fluid antennas\cite{ref110}. Specifically, RISs utilize passive reflection/refraction units deployed between transceivers to intelligently adjust the signal's phase shifts, thereby effectively improving the channel gains. In contrast, movable antennas and fluid antennas dynamically adjust the position or aperture of antennas at the transceivers to create favorable channel conditions. These array architectures have gained widespread attention and proven potential for enhancing system performance.
Nevertheless, existing flexible-antenna systems have limited capability in combating large-scale path loss. For instance, movable and fluid antennas can only adjust their positions within a small spatial range (only a few wavelengths), making them effective mainly for mitigating small-scale fading. While RISs can reconstruct virtual line-of-sight (LoS) links, it inevitably suffers from more severe path loss due to double attenuation \cite{ref111}. 

With the application of high-frequency bands such as millimeter-wave \cite{ref112} and terahertz \cite{ref113}, path loss has become increasingly severe. To overcome this challenge, a new flexible-antenna paradigm known as the pinching antenna (PA) has emerged into the spotlight. In 2021, the first PA prototype was demonstrated by NTT DOCOMO in 60 GHz video transmission system \cite{ref4}. In acknowledgment of this foundational work, we refer to this technology as PASS throughout this paper.
PASS employs a dielectric waveguide to transmit signals. By applying small dielectric particles onto the waveguide, radio waves can be induced and emitted at arbitrary positions along its length. 
Essentially, PASS is a novel implementation of the fluid-antenna and movable-antenna concepts\cite{kitadd1,zhuadd1}, and it resonates with the emerging vision of surface-wave communication superhighways\cite{kitadd2}.
\textit{Why is the pinching antenna chosen?} Compared to the other flexible antennas mentioned above, PASS can also adjust the positions of the PAs to customize the channel, a capability we refer to as \emph{pinching beamforming}\cite{add1}. Additionally, it has the following advantages: 1) \textit{Strong LoS link}: By extending the dielectric waveguide, a PA can be deployed close to the user terminal to form a stable LoS link that effectively reduces large-scale path loss and thus enables a near-wired link. 2) \textit{Scalable deployment}: PASS allows for the flexible formation or termination of communication regions by pinching or uninstalling separate dielectric materials according to communication needs. This is also challenging for existing flexible-antenna systems to achieve.

Given its significant potential, PASS has garnered increasing attention from the academic community. The authors in \cite{ref5} introduced the fundamental architecture of the signal and system model for PASS and analyzed its basic performance gains. In \cite{add11}, the outage probability and average rate in the presence of waveguide losses are analyzed. Further, the authors in \cite{ref6} investigated the array gain of PASS to determine the optimal number of PAs and the optimal antenna spacing. Beyond these fundamental performance analyses, various algorithms have been proposed to further enhance PASS-related wireless transmission \cite{ref7,add2,ref8,ref9,ref10,add12,addkit}. Specifically, a downlink beamforming algorithm based on fractional programming (FP) and block coordinate descent (BCD) was introduced in \cite{ref7} to maximize the weighted sum rate, where each waveguide activates a single PA. This work was further extended a more general case by letting each waveguide pinched with multiple PAs in both uplink and downlink \cite{add2}. A similar problem was also studied in \cite{ref8}, where a penalty-based method is utilized to obtain a stationary-point solution. Additionally, a Transformer-based dual learning algorithm was developed to significantly reduce computational complexity and enhance
performance.  
Another learning-based approach that leveraged graph neural networks (GNNs) was proposed in \cite{ref9}. 
Power minimization and minimum-rate maximization are also important research directions. 
The authors in \cite{ref10} proposed a penalty-based alternating optimization algorithm to minimize transmit power in multiuser downlink scenarios. The authors in \cite{add12} effectively solved the multiuser uplink max-min rate problem by decoupling PAs' positions and resource allocation into two convex subproblems. In \cite{addkit}, a max-min fairness problem in downlink multicast system was considered, where the authors proposed a probability-learning algorithm based on the cross-entropy optimization (CEO) framework to efficiently adjust antenna placement. 

The above research has established an initial foundation for newcomers in this area. However, it is worth noting that most existing works focus on the effectiveness of PASS, particularly in terms of achievable rate. Another critical property in wireless systems, security, has received significantly less research attention. Specifically, due to the broadcast nature of wireless communications, confidential information transmission is highly vulnerable to eavesdropping attacks. To ensure secure transmission, physical layer security  (PLS) techniques can be employed. PLS relies on secure channel coding \cite{ref12}, \cite{ref13} to enhance information security by exploiting the randomness of wireless channels. In PLS, an important performance metric is the secrecy coding rate (or secrecy rate for short), which measures the maximum coding rate at which the transmitter can send confidential information to the legitimate user (Bob) while ensuring that the eavesdropper (Eve) cannot decode it. Recently, secure beamforming has emerged as a promising technology to improve the secrecy rate and further reduce the likelihood of signal interception by Eves \cite{ref103}.

Given the importance of wireless PLS and the superior performance enabled by PASS's pinching beamforming, investigating the application of PASS in PLS is a promising direction. Specifically, the flexible adjustment of the PAs' positions introduces new DoFs to enhance secure beamforming through pinching beamforming. On one hand, the waveguide extension forms a near-wired link near Bob, which limits the signal's propagation range and effectively reduces the risk of information leakage. On the other hand, by customizing a strong LoS channel for Bobs and distancing the PAs from Eves, the capacity of the legitimate channel is increased while the wiretap channel capacity is reduced.

Motivated by these insights, this paper evaluates the gains that PASS brings to the secrecy rate. The main contributions of this work are as follows:
\begin{itemize}
    \item We propose a PASS-enabled framework for enhancing PLS in a multiuser downlink wiretap channel. To evaluate the basic secrecy performance, we derive a closed-form expression for the weighted secrecy sum-rate (WSSR), which is a function of both the baseband beamformer and the pinching beamformer. Building on this, we formulate a joint secure baseband and pinching beamforming design problem to optimize the baseband beamformer and the positions of the PAs, aiming to maximize the WSSR.

    \item We investigate a simplified single-Bob and single-Eve scenario to explore the optimal structure of secure beamforming. Leveraging the matrix determinant lemma, we derive closed-form expressions for the optimal baseband beamformer and the resulting secrecy rate. This allows us to simplify the multivariable optimization problem into a single-variable problem with respect to the pinching beamformer. We then propose a gradient-based method to obtain a stationary-point solution for the optimal positions of the PAs.
    
    \item For the general multiple-Bob and multiple-Eve case, we propose an FP-BCD algorithm to solve the tightly coupled non-convex maximization of the WSSR. Using the Lagrange multiplier method, we derive a closed-form solution for the baseband beamformer. Additionally, we propose a low-complexity Gauss-Seidel approach combined with a one-dimensional search to optimize the pinching beamforming. We prove that this method converges to a stationary-point solution.

	\item We provide extensive numerical results to validate the convergence and effectiveness of the proposed algorithms for joint baseband and pinching beamforming design. The results demonstrate that: i) despite the objective function exhibiting significant cosine oscillations, the proposed algorithms achieve stable and fast convergence, ii) the proposed FP-BCD multiuser secure beamforming design significantly outperforms benchmark schemes using zero-forcing (ZF) and maximal-ratio transmission (MRT), and iii) due to pinching beamforming, PASS achieves a significantly higher secrecy rate than traditional fixed-antenna systems in both single-user and multiuser scenarios.
\end{itemize}

\begin{figure}[!t]
\vspace{-15pt}
    \centering
    \includegraphics[width=0.45\textwidth]{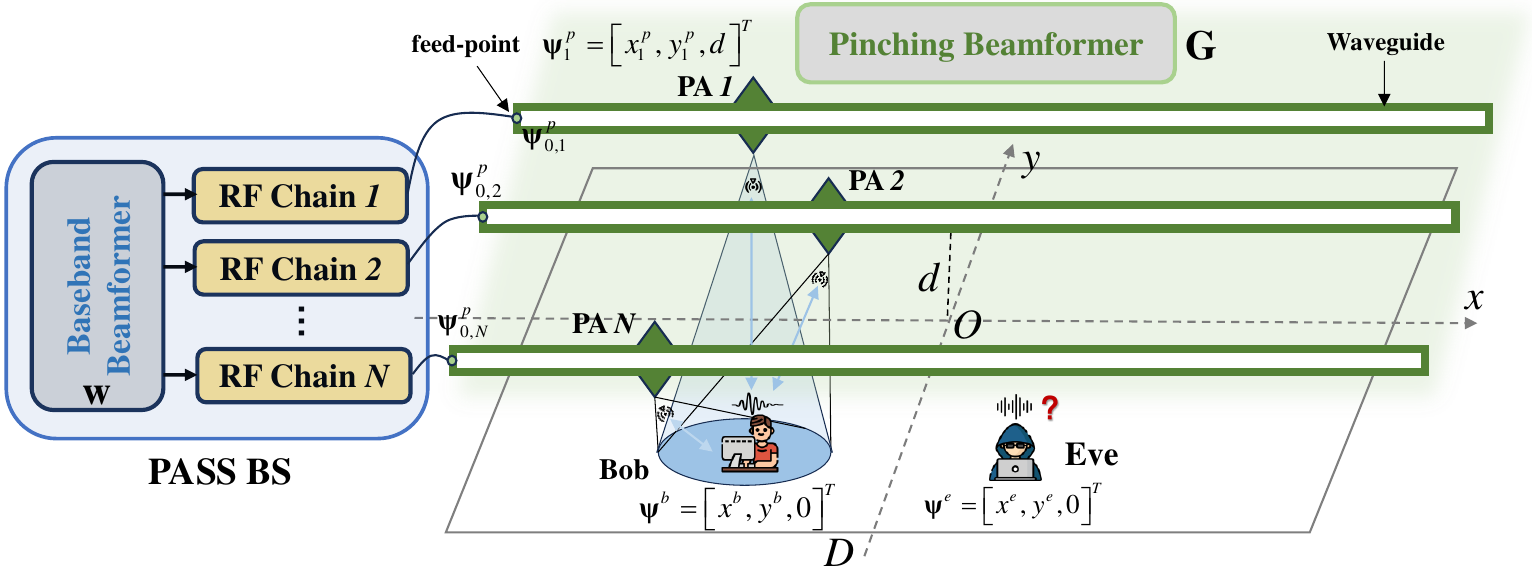} 
    \caption{Illustration of a PASS downlink secure transmission.}
    \label{fig:PA_system} 
\vspace{-20pt}
\end{figure}
The remainder of this paper is organized as follows. Section II introduces the PASS-based secure transmission system and formulates the secrecy rate and WSSR maximization problem. In Section III, an optimal baseband beamformer and a gradient-based algorithm are designed to solve the secrecy rate maximization problem in the single-user case. Section IV presents the FP-BCD algorithm for solving the WSSR maximization problem in the multiuser case. Section V provides numerical results to compare the performance of various approaches under different system configurations. Finally, Section VI concludes the paper.


\section{System Model and Problem Formulation}
A downlink secure transmission system is considered, as shown in Fig. \ref{fig:PA_system}. In PASS, the base station (BS) is equipped with $N$ RF chains, each connected to a waveguide. By applying small dielectric particles, one or more PAs can be activated on each waveguide, with the flexibility to be positioned at any position along the waveguide. The confidential information transmitted by the BS is first processed through baseband beamforming and fed into the RF chains. Subsequently, the positions of the PAs are dynamically adjusted to further evade Eves. Without loss of generality, all waveguides at the BS are assumed to be aligned parallel to the $x$-axis, uniformly spaced, and deployed at a height $d$. All single-antenna Bobs and Eves are distributed within a square area in the $x$-$y$ plane, with a side length of $D$, and its center located at the origin $O=[0,0,0]^T\in \mathbb{R}^{3}$. The length of each waveguide is assumed to be the same as the side length of the region.

\subsection{System Model}
We consider a scenario where $K$ Bobs and $J$ Eves coexist, with multiple PAs deployed on each waveguide. The Cartesian coordinates of Bob $k$ and Eve $j$ are denoted by $\boldsymbol{\psi}^b_k=[x^b_k, y^b_k, 0]^T \in \mathbb{R}^{3}$ and $\boldsymbol{\psi}^e_j=[x^e_j, y^e_j, 0]^T \in \mathbb{R}^{3}$, respectively. 
Let $M_n$ denote the number of PAs on the $n$-th waveguide and the total number of PAs is denoted as $M=\sum_{n=1}^N M_n$. In this case, the transmitted signals at different PAs on the same waveguide are essentially phase-shifted versions of the signal at the feed point. Let $x_k\in \mathbb{C}$ denote the confidential information intended for Bob $k$, which follows a complex Gaussian distribution with zero mean and unit variance, i.e., $x_k\sim \mathcal{CN}(0, 1)$. After applying both baseband beamforming and pinching beamforming, the transmitted signal for Bob $k$ can be expressed as follows:
\begin{align}
\mathbf{s}_k = \underbrace{\mathbf{G}}_{\substack{\text { Pinching } \\
\text { beamforming }}} \cdot \underbrace{\mathbf{w}_k}_{\substack{\text { Baseband} \\
\text { beamforming }}} \cdot x_k.
\label{transmit signals2}
\end{align}
Here, $\mathbf{w}_k \in \mathbb{C}^{N\times 1}$ denotes the baseband beamforming vector for Bob $k$ and $\mathbf{G}$ is given by,
\begin{align}
\mathbf{G} = \begin{bmatrix}
\mathbf{g}_1 & \cdots & \mathbf{0} \\
\vdots & \ddots & \vdots \\
\mathbf{0} & \cdots & \mathbf{g}_N
\end{bmatrix}\in \mathbb{C}^{M \times N},
\end{align}
where
\begin{multline}
\mathbf{g}_n = \frac{1}{\sqrt{M_n}} \left[ e^{-j \frac{2 \pi }{\lambda_p} \left\| \boldsymbol{\psi}_{0,n}^p - \boldsymbol{\psi}_{1,n}^p \right\|}, \ldots, \right. \\
\left. e^{-j \frac{2 \pi }{\lambda_p} \left\| \boldsymbol{\psi}_{0,n}^p - \boldsymbol{\psi}_{M_n,n}^p \right\|} \right]^T \in \mathbb{C}^{M_n \times 1}
\end{multline}
with $\boldsymbol{\psi}_{m,n}^p=[x_{m,n}^p,y_n^p,d]^T \in \mathbb{R}^{3}$ denoting the position of $m$th PA on the $n$th waveguide, $\boldsymbol{\psi}_{0,n}^p\in \mathbb{R}^{3}$ indicating the position of the feed point of the $n$th waveguide, and $\lambda_p$ representing the waveguide wavelength in a dielectric medium.

The legitimate channels between all PAs and Bob $k$ are denoted as $\hat{\mathbf{h}}_k^b=[\hat{\mathbf{h}}_{1,k}^b,\ldots,\hat{\mathbf{h}}_{N,k}^b]^T\in \mathbb{C}^{M\times 1}, \forall k \in \{1, \dots, K\}$, and the wiretap channels from all PAs to Eve $j$ are denoted as $\hat{\mathbf{h}}_j^e=[\hat{\mathbf{h}}_{1,j}^e,\ldots,\hat{\mathbf{h}}_{N,j}^e]^T\in \mathbb{C}^{M\times 1}, \forall j \in \{1, \dots, J\}$, where $\hat{\mathbf{h}}_{n,k}^b \in \mathbb{C}^{1\times M_n}$ and $\hat{\mathbf{h}}_{n,k}^e\in \mathbb{C}^{1\times M_n}$ represent the channels from the $n$th waveguide to Bob $k$ and Eve $j$, respectively.
The $m$th element of the channels $\hat{\mathbf{h}}_{n,k}^b$ and $\hat{\mathbf{h}}_{n,j}^e$ can be modeled using the following free-space path loss model\footnote{In this paper, only LoS links are considered, as it is typically over 20 dB stronger than non-line-of-sight (NLoS) paths\cite{LoS}. The extension to NLoS scenarios will be explored in future work.}:
\begin{subequations}
\begin{align}\label{channel_multiBobEve}
[\hat{\mathbf{h}}_{n,k}^b]_m=\frac{\sqrt{\eta} e^{-j \frac{2 \pi }{\lambda_c} \left\| \boldsymbol{\psi}^b_k - \boldsymbol{\psi}^p_{m,n}\right\|}}{\left\| \boldsymbol{\psi}^b_k - \boldsymbol{\psi}^p_{m,n} \right\|},\\
[\hat{\mathbf{h}}_{n,j}^e]_m=\frac{\sqrt{\eta} e^{-j \frac{2 \pi }{\lambda_c} \left\| \boldsymbol{\psi}^e_j - \boldsymbol{\psi}^p_{m,n}\right\|}}{\left\| \boldsymbol{\psi}^e_j - \boldsymbol{\psi}^p_{m,n} \right\|},
\end{align}
\end{subequations}
where $\eta=\frac{\lambda_c^2}{16\pi^2}$ is a constant with $\lambda_c$ denoting the carrier wavelength in the free space.  
The channels are assumed to be quasi-static block-fading, where the channel remains constant within each fading block. Due to the reconfigurable positions of the PAs, the channels can be reconfigured. 

The received signal at Bob $k$ can then be expressed as follows:
\begin{align}\label{single_PA_bob}
y_k^b = \hat{\mathbf{h}}_k^{bT}\mathbf{G}\mathbf{w}_k x_k+\sum_{i=1,i \neq k}^K\hat{\mathbf{h}}_k^{bT}\mathbf{G}\mathbf{w}_i x_i +n_k^b,
\end{align}
where $n_k^b\sim \mathcal{CN}(0, \sigma_{b,k}^{2})$ is the additive white Gaussian noise at Bob $k$, with $\sigma_{b,k}^{2}$ denoting the noise power.
Similarly, the signal overheard by Eve $j$ can be represented as
\begin{align} \label{single_PA_eve}
y_j^e = \sum_{i=1}^K\hat{\mathbf{h}}_j^{eT}\mathbf{G}\mathbf{w}_i x_i+n_j^e, 
\end{align}
where $n_j^e\sim \mathcal{CN}(0, \sigma_{e,j}^{2})$ denotes the noise of Eve $j$. Since the channels $\hat{\mathbf{h}}_k^b$, $\hat{\mathbf{h}}_j^e$ and pinching beamforming matrix $\mathbf{G}$ all depend on $\mathbf{x}^p$, we define for simplicity: ${\mathbf{h}}_k^b(\mathbf{x}^p)= \mathbf{G}^T\hat{\mathbf{h}}_k^b=[\mathbf{g}_1^T \hat{\mathbf{h}}_{1,k}^{bT},\ldots,\mathbf{g}_N^T \hat{\mathbf{h}}_{N,k}^{bT}]^T \in \mathbb{C}^{N\times1}$, ${\mathbf{h}}_j^e(\mathbf{x}^p) = \mathbf{G}^T\hat{\mathbf{h}}_j^e=[\mathbf{g}_1^T \hat{\mathbf{h}}_{1,j}^{eT},\ldots,\mathbf{g}_N^T \hat{\mathbf{h}}_{N,j}^{eT}]^T\in \mathbb{C}^{N\times1}$, $\forall k \in\{1,\ldots,K\},~\forall j \in\{1,\ldots,J\}$, where $\mathbf{x}^p = [\mathbf{x}_1^p, \ldots, \mathbf{x}_N^p]^T \in \mathbb{R}^{M\times1}$ represents the set of all PA coordinates along the $x$-axis with $\mathbf{x}_n^p = [x_{1,n}^p,\ldots,x_{M_n,n}^p]$.

\subsection{Problem Formulation}
Assume that Eves act as a disguised legitimate user and transmits uplink pilot signals, the BS allows them to receive the common message while suppressing the confidential message. This is a widely adopted assumption in broadcast scenarios with confidential communications, as also considered in \cite{assu1}, \cite{assu2} and the references therein. Under this assumption, the BS can obtain the channel state information (CSI) of both Bobs and Eves.
The signal-to-interference-plus-noise ratio (SINR) of the confidential signal received by Bob $k$ is given by
\begin{align}
{\gamma}_k = \frac{|\mathbf{h}_k^{bT}(\mathbf{x}^p) \mathbf{w}_k|^2}{\sum_{i \neq k} |\mathbf{h}_k^{bT}(\mathbf{x}^p) \mathbf{w}_i|^2 + \sigma_{b,k}^2}.
\end{align}
In a multi-Bob and multi-Eve scenario, evaluating the information leakage rate is non-trivial. From a worst-case perspective, all Eves are assumed to have perfect CSI knowledge and cooperate to cancel interference from other Bobs while performing maximal-ratio combining. This helps characterize the worst-case leakage performance and ensures that confidential messages can be transmitted reliably under conservative assumptions. Under this model, the aggregated signal-to-noise ratio (SNR) for eavesdropping Bob $k$' signal can be expressed as follows:
\begin{align}
{\Gamma}_{k} = \sum_{j=1}^{J}\frac{|\mathbf{h}_j^{eT}(\mathbf{x}^p) \mathbf{w}_k|^2}{\sigma_{e,j}^2}.
\end{align}
Therefore, the secrecy rate of Bob $k$ is given by $R_k^{\rm sec}=\max\big\{\log_2\left(\frac{1 + {\gamma}_k}{1 + {\Gamma}_{k}}\right),0\big\}$. To evaluate the overall secrecy performance of the system, the WSSR of the system is defined as $R^{\rm sec}= \sum_{k=1}^{K}\alpha_k R_k^{\rm sec}$, where $\alpha_k>0, \forall k \in\{1,\ldots,K\}$, denotes the priority weight of Bob $k$. 
It can be observed from $R^{sec}$ that compared to traditional fixed-position antennas, the additional spatial DoFs introduced by PAs enable not only the design of the baseband beamformer but also the reconfiguration of the channels. This allows for a joint optimization  of $\{\mathbf{w}_k\}_{k=1}^{K}$ and $\mathbf{x}^p$ to maximize the WSSR. Therefore, the joint optimization problem can be formulated as follows:
\begin{subequations}
\begin{align}\label{problem_multiple_LUTEVE}
\mathcal{P}_1: ~&\max_{\{\mathbf{w}_k\}_{k=1}^{K},\mathbf{x}^p}~R^{\rm sec}\\
~~& {\rm{s.t.}}~\sum_{k=1}^{K}\|\mathbf{w}_k\|^2\leq P_T,\label{C1}\\
~~& x_{m,n}^p \in \left[-D/2,D/2\right],\forall n=1,\ldots,N,\label{C2}\\
~~& |x_{m,n}^p-x_{m',n}^p|>\Delta_{\rm min},\forall m \neq m',\forall n,\label{C3}
\end{align}
\end{subequations}
where \eqref{C1} denotes the maximum transmit power of the BS, \eqref{C2} and \eqref{C3} ensure that the PAs' positions do not exceed the waveguide length and prevent mutual coupling between different PAs on the same waveguide\cite{ref15}, respectively.
Observing problem $\mathcal{P}_1$, it presents three analytical challenges:  
1) Non-differentiability: the presence of the $\max\{\cdot\}$ operation makes the objective function non-differentiable.  
2) Non-convexity: the objective function is inherently non-convex.  
3) Strong Coupling: the optimization variables are strongly coupled, adding complexity to the problem.  
In the sequel, we aim to design an efficient algorithm to tackle problem $\mathcal{P}_1$.

\section{Joint Secure Beamforming for Single-User Scenario}
In this section, we consider a basic case of problem $\mathcal{P}_1$ with a single Bob and a single Eve, where one PA is activated on each waveguide, to explore the structure of the optimal baseband beamforming.
First, a simplified secrecy rate maximization problem is formulated. Then, the optimal baseband beamforming vector is derived. Finally, a gradient ascent-based method is proposed to obtain a locally optimal solution for the PAs' positions. 

\subsection{Problem Formulation}
Assume that only one PA is activated on each waveguide of the BS. The position of the PA on the $n$-th waveguide is denoted as $\boldsymbol{\psi}_n^p=[x_n^p,y_n^p,d]^T \in \mathbb{R}^{3}$. For convenience, the collection of the $x$-axis coordinates of the $N$ PAs is denoted as $\mathbf{x}^p = [x_1^p, \dots, x_N^p]^T$. The Cartesian coordinates of Bob and Eve are given by $\boldsymbol{\psi}^b=[x^b, y^b, 0]^T \in \mathbb{R}^{3}$ and $\boldsymbol{\psi}^e=[x^e, y^e, 0]^T \in \mathbb{R}^{3}$. Let the confidential information sent to Bob be $x^b\sim \mathcal{CN}(0, 1)\in \mathbb{C}$. Then, the transmitted signal from the BS can be written as follows:
\begin{align}
\mathbf{s}^b = \mathbf{G} \cdot \mathbf{w} \cdot x^b.
\label{transmit signals}
\end{align}
Here, $\mathbf{w}\in\mathbb{C}^{N}$ is the baseband beamforming vector for Bob, and $\mathbf{G}=\text{diag}\left\{e^{-j \frac{2 \pi }{\lambda_p}\left\| \boldsymbol{\psi}_{0,1}^p - \boldsymbol{\psi}_1^p\right\|},\ldots,e^{-j \frac{2 \pi }{\lambda_p}\left\| \boldsymbol{\psi}_{0,N}^p - \boldsymbol{\psi}_N^p\right\|}\right\}\in \mathbb{C}^{N\times N}$ denotes the pinching beamforming matrix.

The BS-to-Bob and BS-to-Eve channels can be represented as $\hat{\mathbf{h}}^b=\left[\hat{h}_{1}^b,\ldots,\hat{h}_{N}^b\right]^T \in\mathbb{C}^{N\times1}$ and $\hat{\mathbf{h}}^e=\left[\hat{h}_{1}^e,\ldots,\hat{h}_{N}^e\right]^T\in\mathbb{C}^{N\times1}$, respectively. Here, $\hat{h}_{n}^i$ denotes the channel from the $n$th PA to either Bob or Eve, which is given by 
\begin{align}\label{channel}
\hat{h}_{n}^i=\frac{\sqrt{\eta} e^{-j \frac{2 \pi}{\lambda_c} \left\| \boldsymbol{\psi}^i - \boldsymbol{\psi}_n^p\right\|}}{\left\| \boldsymbol{\psi}^i - \boldsymbol{\psi}_n^p \right\|}, ~~i \in\{b,e\}.
\end{align}
Therefore, the received signals at Bob and Eve can be written as follows:
\begin{align}
y^i= \hat{\mathbf{h}}^{iT}\mathbf{s}^b + n^i, ~~i \in\{b,e\},
\label{received signals}
\end{align}
where $n^i\sim \mathcal{CN}(0, \sigma_i^2)$ is the additive white Gaussian noise with $\sigma_i^2$ denoting the noise power.

The secrecy rate can be defined as follows\cite{ref16}:
\begin{align}\label{secrecy rate}
R^{\rm sec} = \max\Bigg\{\log_2\left(\frac{1 + \left|\hat{\mathbf{h}}^{bT}\mathbf{G}\mathbf{w}\right|^2/\sigma_b^2}{1 + \left|\hat{\mathbf{h}}^{eT}\mathbf{G}\mathbf{w}\right|^2/\sigma_e^2}\right),0\Bigg\}.
\end{align}
For simplicity, let ${\mathbf{h}}^i(\mathbf{x}^p) = \mathbf{G}\hat{\mathbf{h}}^i=[{h}_{1}^i,\ldots,{h}_{N}^i]^T \in \mathbb{C}^{N\times 1}$, with ${h}_n^i=\hat{h}_n^i e^{-j \frac{2 \pi }{\lambda_p}\left\| \boldsymbol{\psi}_{0,n}^p - \boldsymbol{\psi}_n^p\right\|}$. Equation \eqref{secrecy rate} can be simplified as follows:
\begin{align}\label{secrecy rate1}
R^{\rm sec} = \max\bigg\{\log_2\left(\frac{1 + \left|\mathbf{h}^{bT}(\mathbf{x}^p)\mathbf{w}\right|^2/\sigma_b^2}{1 + \left|\mathbf{h}^{eT}(\mathbf{x}^p)\mathbf{w}\right|^2/\sigma_e^2}\right),0\bigg\}.
\end{align}

Accordingly, we formulate the following optimization problem, where the joint optimization of the baseband beamforming vector $\mathbf{w}$ and the positions $\mathbf{x}^p$ of the PAs is performed to maximize the secrecy rate:
\begin{subequations}\label{problem_single_LUTEVE}
\begin{align}
\mathcal{P}_2: ~&\max_{\mathbf{w},\mathbf{x}^p}~R^{\rm sec}\\
~~ &{\rm{s.t.}}~\left|\left|\mathbf{w}\right|\right|^2\leq P_T,\\
~~ &x_n^p \in \left[-D/2,D/2\right],\forall n=1,\ldots,N\label{singleC2}, 
\end{align}
\end{subequations}
where $P_T$ is the power budget. 

\subsection{The Optimal Baseband Digital Beamformer Design}
Due to the non-convex nature of the objective function and the strong coupling between the optimization variables $\mathbf{w}$ and $\mathbf{x}^p$, it is challenging to solve.
To address this, we first design $\mathbf{w}$ while keeping $\mathbf{x}^p$ fixed. For simplicity of notation, the $(\mathbf{x}^p)$ notation is omitted in the following analysis.
By referring to problem $\mathcal{P}_2$ defined in \eqref{problem_single_LUTEVE}, the baseband beamformer $\mathbf{w}^\star$ that maximizes the secrecy rate is given by
\begin{equation}\label{w_star_initial}
\mathbf{w}^\star = \argmax_{\|\mathbf{w}\|^2 \leq P_T} \frac{1 + \left|\mathbf{h}^{bT}\mathbf{w}\right|^2/\sigma_b^2}{1 + \left|\mathbf{h}^{eT}\mathbf{w}\right|^2/\sigma_e^2}.
\end{equation}

Based on the monotonicity of the function $f(x) = \frac{1 + ax}{1 + bx}$ for $x > 0$, $a >b> 0$, it can be easily proven that \eqref{w_star_initial} achieves its maximum when $\|\mathbf{w}\|^2 = P_T$. Let $\mathbf{w}=\sqrt{P_T}\mathbf{v}$ and $\|\mathbf{v}\|^2 = 1$, the problem in \eqref{w_star_initial} can be equivalently rewritten as follows:
\begin{equation}\label{v_star_initial}
\mathbf{v}^\star = \argmax_{\|\mathbf{v}\|^2 = 1} \frac{1 + |\mathbf{h}^{bT} \mathbf{v}|^2 \hat{\gamma}_b}{1 + |\mathbf{h}^{eT} \mathbf{v}|^2 \hat{\gamma}_e}.
\end{equation}
where $\hat{\gamma}_i\triangleq P_T/\sigma_i^{2}$ for $i\in\{b,e\}$.
The problem in \eqref{v_star_initial} can be further expressed in the form of the following Rayleigh quotient \cite{ref17}:
\begin{equation}\label{v_Rayleigh quotient}
\mathbf{v}^\star = \argmax_{\|\mathbf{v}\|^2 = 1} \frac{\mathbf{v}^H (\mathbf{I}+\hat{\gamma}_b\mathbf{h}^{b*}\mathbf{h}^{bT})\mathbf{v}}{\mathbf{v}^H (\mathbf{I}+\hat{\gamma}_e\mathbf{h}^{e*}\mathbf{h}^{eT})\mathbf{v}}.
\end{equation}
Based on \cite{ref17}, the optimal solution to problem \eqref{v_Rayleigh quotient} is given by $\mathbf{v}^\star=\frac{(\mathbf{I}+\hat{\gamma}_e\mathbf{h}^{e*}\mathbf{h}^{eT})^{-1/2}\mathbf{p}}{\|(\mathbf{I}+\hat{\gamma}_e\mathbf{h}^{e*}\mathbf{h}^{eT})^{-1/2}\mathbf{p}\|^2}$, where $\mathbf{p}$ denotes the principal eigenvector of the matrix $\mathbf{\Delta}=(\mathbf{I}+\hat{\gamma}_e\mathbf{h}^{e*}\mathbf{h}^{eT})^{-1/2}(\mathbf{I}+\hat{\gamma}_b\mathbf{h}^{b*}\mathbf{h}^{bT})(\mathbf{I}+\hat{\gamma}_e\mathbf{h}^{e*}\mathbf{h}^{eT})^{-1/2}$. The resulting maximum secrecy rate is given by $R^{\rm sec}=\log_2\mu_{\mathbf{\Delta}}$, where $\mu_{\mathbf{\Delta}}$ is the principal eigenvalue of $\mathbf{\Delta}$.
By further using the matrix determinant lemma, we derive a closed-form expression for $R^{\rm sec}$, which is presented in the following lemma.

\begin{lemma}\label{Lemma_for_mu_Delta}
Given the channel $\mathbf{h}^i$, $~i\in\{b,e\}$ and the transmit power $P_T$, the maximum secrecy rate is given by
\begin{align}\label{optimal_mu_Delta}
R^{\rm sec}=\log_2\left( 1 + \frac{b + \sqrt{b^2 + 4ac}}{2a}\right),
\end{align}
where $a=1+\hat{\gamma}_e \|\mathbf{h}^e\|^2$, $b=\hat{\gamma}_b \|\mathbf{h}^b\|^2-\hat{\gamma}_e \|\mathbf{h}^e\|^2+\hat{\gamma}_b\hat{\gamma}_e \|\mathbf{h}^b\|^2\|\mathbf{h}^e\|^2-\hat{\gamma}_b\hat{\gamma}_e |\mathbf{h}^{bT}\mathbf{h}^{e*}|^2$, and $c=\hat{\gamma}_b\hat{\gamma}_e \|\mathbf{h}^b\|^2\|\mathbf{h}^e\|^2  - \hat{\gamma}_b\hat{\gamma}_e |\mathbf{h}^{bT}\mathbf{h}^{e*}|^2$.
\end{lemma}
\vspace{-5pt}

\begin{IEEEproof}
Please refer to Appendix~\ref{Appendix_A} for more details.
\end{IEEEproof}

\begin{remark}
The results in \textbf{Lemma~\ref{Lemma_for_mu_Delta}} indicate that, for a given positions $\mathbf{x}^p$ of the PAs, the optimal secrecy rate can be achieved when the baseband beamformer is $\mathbf{w}^{\star}=\sqrt{P_T}\mathbf{v}^{\star}$. It is worth noting that the optimal secrecy rate $R^{sec}$ is a function of the positions $\mathbf{x}^p$. Therefore, the problem in \eqref{problem_single_LUTEVE} can be equivalently reformulated as follows:
\begin{align}\label{problem_for_T}
\mathcal{P}_{2-1}: ~~&\max_{\mathbf{x}^p}~f(\mathbf{x}^p)=\frac{b(\mathbf{x}^p) + \sqrt{b(\mathbf{x}^p)^2 + 4a(\mathbf{x}^p)c(\mathbf{x}^p)}}{a(\mathbf{x}^p)},\\
~~ &{\rm{s.t.}}~ \eqref{singleC2}.\nonumber
\end{align}
\end{remark}

\subsection{The Gradient-Based Method for Optimizing Pinching Beamforming}
We propose an efficient method to optimize the PAs' positions $\mathbf{x}^p$. Due to the coupling between $\{x_n^p\}_{n=1}^N$, solving the problem $\mathcal{P}_{2-1}$ is a challenging task.
To this end, we propose a elementwise-BCD approach. Specifically, the problem is decomposed into $N$ subproblems, where each subproblem optimizes the position $x_n^p$ of the $n$-th PA while fixing the positions of all other PAs. By iteratively solving these $N$ subproblems, a locally optimal solution of problem $\mathcal{P}_{2-1}$ can be obtained.

\begin{algorithm}[!t]
\caption{Gradient Ascent-Based Algorithm for Solving $\mathcal{P}_{2-2}$}\label{alg1}
\begin{algorithmic}[1]
\State Initialize $(\mathbf{x}^p)^0 = [(x_1^p)^0, \ldots, (x_N^p)^0]$, the maximum iteration number $I$, step size $\beta_{\text{ini}}$, the minimum tolerance step size $\beta_{\text{min}}$, and set the current iteration $t = 0$;
\Repeat
    \For{$n = 1$ to $N$}
        \State Compute the gradient value $\nabla_{(x_n^p)^{t}} f((x_n^p)^{t})$ and set $\beta_t = \beta_{\text{ini}}$;
        \Repeat
            \State Compute $\hat{x}_n^p = (x_n^p)^{t} + \beta_t \cdot \nabla_{(x_n^p)^{t}} f((x_n^p)^{t})$ and set $\beta_t = \beta_t / 2$;
        \Until{$\hat{x}_n^p \in [-D/2, D/2]$ \textbf{and} $f(\hat{x}_n^p) > f((x_n^p)^{t})$ \textbf{or} $\beta_t < \beta_{\text{min}}$}
        \State Set $(x_n^p)^{t} = \hat{x}_n^p$ and update $(x_n^p)^{t+1} = \hat{x}_n^p$;
    \EndFor
    \State Update $t = t + 1$;
\Until{convergence or the maximum iteration number $I$ is reached;}
\State Update $\mathbf{w}$ by $\mathbf{w}^* = \sqrt{P_T} \mathbf{v}^*$.
\end{algorithmic}
\end{algorithm}

Subsequently, we focus on optimizing $x_n^p$ while keeping $\{x_{n'}^p\}_{n' \neq n}$ fixed. The corresponding subproblem is formulated as follows:
\begin{subequations}
\begin{align}\label{problem_for_T1}
\mathcal{P}_{2-2}: ~~&\max_{x_n^p} f(x_n^p)=\frac{b(x_n^p) + \sqrt{b(x_n^p)^2 + 4a(x_n^p)c(x_n^p)}}{a(x_n^p)},\\
~~ &{\rm{s.t.}}~ x_n^p \in \left[-D/2,D/2\right].  
\end{align}
\end{subequations}
To solve problem $\mathcal{P}_{1-2}$, we can use a gradient-based method. The updating rule can be formulated as $(x_n^p)^{t+1}= (x_n^p)^{t}+\beta_t\nabla_{(x_n^p)^{t}} f((x_n^p)^{t})$, where $\beta_t$ is the step size in the $t$-th iteration. The derivative of $f(x_n^p)$ with respect to $x_n^p$ can be found in \eqref{dfdx} given in Appendix~\ref{Appendix_B}.

The overall algorithm for solving $\mathcal{P}_{2-2}$ is outlined in \textbf{Algorithm \ref{alg1}}, which is guaranteed to converge to a stationary-point solution \cite{add3}.
Due to the presence of exponential terms in the channel function, the objective function in \eqref{problem_for_T1} exhibits numerous local maxima. However, experimental results reveal that even in the presence of cosine oscillations in the objective function, the gradient-based method remains stable and converges after multiple iterations. By selecting an appropriate step size, the monotonic increase of the objective function value can be effectively ensured. 
To visually illustrate the convergence process of the gradient-based method, we introduce the average norm of gradient as a visualization metric, which is defined as $\nabla_{\rm norm}=\frac{1}{N}\sum_{n=1}^N|\nabla_{x_n^p} f(x_n^p)|$. The convergence curve of \textbf{Algorithm \ref{alg1}} with respect to the secrecy rate and gradient norm can be seen in Fig. \ref{fig:conver} of Section V. Reviewing \textbf{Algorithm \ref{alg1}}, its primary computational complexity arises from the matrix inversion operation in $\mathbf{v}^{\star}$ and the iterative process of gradient ascent. The complexity scales as $\mathcal{O}(N^3 + I N \log_2 \frac{1}{\beta_{\min}})$.

\section{Joint Secure Beamforming for Multi-User Scenario}
For the multi-user case, we are unable to find the optimal baseband beamforming solution.
In this section, we will present a FP-BCD algorithm to effectively address the secure transmission problem in this case. First, the closed-form solution for the baseband beamforming vector is derived using the Lagrange multiplier method. Then, the pinching beamformer is optimized using a Gauss-Seidel-based search approach.

\subsection{FP-BCD Algorithm}
Observing problem $\mathcal{P}_1$, we streamline the notation by normalizing each channel with its corresponding noise power, i.e., $\mathbf{h}_k^b(\mathbf{x}^p)=\frac{\mathbf{h}_k^b(\mathbf{x}^p)}{\sigma_{b,k}}$ and $\mathbf{h}_j^e(\mathbf{x}^p)=\frac{\mathbf{h}_j^e(\mathbf{x}^p)}{\sigma_{e,j}}$. Based on this, \eqref{problem_multiple_LUTEVE} can be rewritten as \eqref{problem_multiple_LUTEVE0} at the bottom of this page.
\begin{figure*}[hb] 
\vspace{-5pt}
\centering 
\hrule 
\vspace{-5pt}
\begin{subequations}\label{problem_multiple_LUTEVE0}
\begin{align}
\max_{\{\mathbf{w}_k\}_{k=1}^{K},\mathbf{x}^p}~&\sum_{k=1}^{K}\alpha_k\max\Big\{\log_2\left(1+\frac{\left|\mathbf{h}_k^{bT}(\mathbf{x}^p) \mathbf{w}_k\right|^2}{\sum_{i=1, i \neq k}^K\left|\mathbf{h}_k^{bT}(\mathbf{x}^p) \mathbf{w}_i\right|^2+1}\right)
-\log_2\left(1+\sum_{j=1}^{J}\left\lvert\mathbf{h}_j^{eT}(\mathbf{x}^p) \mathbf{w}_k\right\rvert^2\right),0\Big\}\label{problem_multiple_LUTEVE_transformation0_Objective}\\
&{\rm{s.t.}}~\eqref{C1},\eqref{C2},\eqref{C3}.\nonumber
\end{align}
\end{subequations}
\vspace{-20pt}
\end{figure*}
The operator $\max\{x,0\}$ in \eqref{problem_multiple_LUTEVE_transformation0_Objective} makes the problem \eqref{problem_multiple_LUTEVE0} difficult to solve. We address this by replacing it with its equivalent form using the following lemma.
\begin{lemma}\label{Lemma_MAX}
Given the legitimate channels $\{\mathbf{h}_k^b\}_{k=1}^{K}$ and wiretap channels $\{\mathbf{h}_j^e\}_{j=1}^{J}$, problem \eqref{problem_multiple_LUTEVE0} can be reformulated as the following equivalent problem:
\begin{subequations}\label{transformation1}
\begin{align}
\max_{\substack{\{\mathbf{w}_k\}_{k=1}^{K}, \mathbf{x}^p, \\ \{\tau_k\}_{k=1}^{K}}}~&\sum_{k=1}^{K}\tau_k(\log_2\left(1+\gamma_k\right)
-\log_2\left(1+\Gamma_k\right))\label{current_transformation1_Objective}\\
&{\rm{s.t.}}~\eqref{C1},\eqref{C2},\eqref{C3},\nonumber\\
&\tau_k\in[0,\alpha_k], \forall k\label{C4}.
\end{align}
\end{subequations}
\end{lemma}

\begin{IEEEproof}
This is proven by showing that problem \eqref{transformation1} can be equivalently transformed into \eqref{problem_multiple_LUTEVE0}. It can be easily seen that, given $\{\mathbf{w}_k\}_{k=1}^{K}, \mathbf{x}^p$, the optimal $\tau_k$ satisfies
\begin{align}\label{bk}
\tau_k^{\star}=\begin{cases}
\alpha_k& \gamma_k\geq \Gamma_k,\\
0& \gamma_k< \Gamma_k.
\end{cases}
\end{align}
Substituting $\tau_k=\tau_k^{\star}$ into \eqref{current_transformation1_Objective} establishes the equivalence between \eqref{transformation1} and \eqref{problem_multiple_LUTEVE0}.
\end{IEEEproof}

Similarly, due to the non-convexity of the objective function and the coupling between the optimization variables, we will use a BCD-based method to faciliate the optimization \cite{ref18}. To simplify the expressions, before solving for the PAs' positions, the channels are considered constant, and we omit the notation of $(\mathbf{x}^p)$.
Inspired by the FP framework\cite{ref121}, we present the following lemma to further transform it into an equivalent form that is more tractable. 

\begin{lemma}\label{Lemma_FP_Must}
Given $\{\mathbf{h}_k^b\}_{k=1}^{K}$ and $\{\mathbf{h}_j^e\}_{j=1}^{J}$, problem \eqref{transformation1} is equivalent to the following:
\begin{subequations}\label{transformation2}
\begin{align}
\max_{\substack{\{\mathbf{w}_k\}_{k=1}^{K}, \mathbf{x}^p, \\ \{\tau_k\}_{k=1}^{K}}}~&\sum_{k=1}^{K}\tau_k\Bigg(\log_2\left(1+\gamma_k\right)
+\log_2\left(1+\frac{G_{\Gamma}-\Gamma_k}{1+\Gamma_k}\right)\nonumber\\
&\quad\quad\quad\quad-\log_2\left(1+G_{\Gamma}\right)\Bigg)\label{transformation2_Objective}\\
&{\rm{s.t.}}~\eqref{C1},\eqref{C2},\eqref{C3},\eqref{C4}.\nonumber
\end{align}
\end{subequations}
where $G_{\Gamma}\triangleq P_T\sum_{j=1}^{J}\| \mathbf{h}_j^e\|^2\geq \Gamma_k$.
\end{lemma}
\vspace{-5pt}
\begin{IEEEproof}
According to the Cauchy-Schwarz inequality, we have
\begin{subequations}
\begin{align}
\Gamma_{k}=\sum_{j=1}^{J}{\left\lvert \mathbf{h}_j^{eT} \mathbf{w}_k\right\rvert^2}&\leq \sum_{j=1}^{J}\| \mathbf{h}_j^e\|^2\| \mathbf{w}_{k}\|^2 \nonumber\\
&\leq P_T\sum_{j=1}^{J}\|\mathbf{h}_j^e\|^2=G_{\Gamma},
\end{align}
\end{subequations}
which yields $G_{\Gamma}-\Gamma_{k}\geq 0$. It follows that
\begin{subequations}
\begin{align}
-\log_2\left({1+\Gamma_{k}}\right)&=\log_2\left(\frac{1+G_{\Gamma}}{1+\Gamma_{k}}\right)-\log_2\left(1+G_{\Gamma}\right)\nonumber\\
&=\log_2\left(1+\frac{G_{\Gamma}-\Gamma_{k}}{1+\Gamma_{k}}\right)-\log_2\left(1+G_{\Gamma}\right).
\end{align}
\end{subequations}
Therefore, the equivalence between \eqref{transformation1} and \eqref{transformation2} can be established. The non-negativity of the numerator in $\frac{G_{\Gamma}-\Gamma_{k}}{1+\Gamma_{k}}$ ensures that \eqref{transformation2_Objective} satisfies the condition for employing the Lagrange dual transform\cite{ref121}.
\end{IEEEproof}
We next use the following lemma to equivalently transform problem \eqref{transformation2} to a more tractable form.
\begin{lemma}\label{Lemma_FP_Must}
The Lagrange dual transform can be applied to transform problem \eqref{transformation2} to a equivalent problem \eqref{transformation3}, shown at the bottom of next page. The optimal $\mu_k$ and $\nu_k$ are given by $\mu_k^{\star}=\gamma_k$ and $\nu_k^{\star}=\frac{G_{\Gamma}-\Gamma_k}{1+\Gamma_k}$, respectively.
\end{lemma}
\begin{figure*}[hb] 
\vspace{-15pt}
\centering 
\hrule 
\begin{subequations}\label{transformation3}
\begin{align}
&\max_{\mathbf{w}_k,\mathbf{x}^p,\tau_k,\mu_k,\nu_k}\sum_{k=1}^{K}\tau_k\left(\log(1+\mu_k)\!-\!\mu_k\!+\!
\frac{(1+\mu_k)\left\lvert \mathbf{h}_k^{bT} \mathbf{w}_k\right\rvert^2}
{1+\sum_{i=1}^{K}\left\lvert \mathbf{h}_k^{bT} \mathbf{w}_i\right\rvert^2}\!+\!
\log(1+\nu_k)\!-\!\nu_k\!+\!(1\!+\!\nu_k)\frac{G_{\Gamma}-\Gamma_k}{1+G_{\Gamma}}\!-\!\log_2\left(1\!+\!G_{\Gamma}\right)\right)\label{transformation3_Objective}\\
&{\rm{s.t.}}~\eqref{C1},\eqref{C2},\eqref{C3},\eqref{C4}.\nonumber
\end{align}
\end{subequations}
\vspace{-15pt}
\hrule 
\end{figure*}

\begin{IEEEproof}
Upon examining \eqref{transformation3}, we observe that when $\{\mathbf{w}_k\}_{k=1}^{K}$, $\mathbf{x}^p$, $\{\tau_k\}_{k=1}^{K}$, and $\{\nu_k\}_{k=1}^{K}$ are fixed, the problem becomes a convex optimization problem with respect to $\mu_k$, which can be solved using the first-order optimality conditions, i.e.,
\begin{align}
\frac{\partial}{\partial \mu_k}\eqref{transformation3_Objective}&=\frac{1}{1+\mu_k}-1+\frac{\left\lvert\mathbf{h}_k^{bT} \mathbf{w}_k\right\rvert^2}
{1+\sum_{i=1}^{K}\left\lvert\mathbf{h}_k^{bT} \mathbf{w}_i\right\rvert^2}=0.
\end{align}
We can easily obtain the optimal $\mu_k^{\star}=\frac{\left\lvert \mathbf{h}_k^{bT} \mathbf{w}_k\right\rvert^2}
{1+\sum_{i=1,i\ne k}^{K}\left\lvert\mathbf{h}_k^{bT} \mathbf{w}_i\right\rvert^2}=\gamma_k$. Similarly, for solving $\nu_k$,
\begin{align}
\frac{\partial}{\partial \nu_k}\eqref{transformation3_Objective}&=\frac{1}{1+\nu_k}-1+\frac{G_{\Gamma}-\Gamma_k}{1+G_{\Gamma}}=0,
\end{align}
which yields the optimal $\nu_k^{\star}=\frac{G_{\Gamma}-\Gamma_k}{1+\Gamma_k}$.
Substituting $\mu_k^{\star}$ and $\nu_k^{\star}$ back into \eqref{transformation3_Objective} retrieves the objective function in \eqref{transformation2_Objective}, thereby confirming the equivalence of these two problems.
\end{IEEEproof}

Due to the presence of the fractional term $\frac{(1+\mu_k)\left\lvert\mathbf{h}_k^{bT} \mathbf{w}_k\right\rvert^2}{1+\sum_{i=1}^{K}\left\lvert\mathbf{h}_k^{bT} \mathbf{w}_i\right\rvert^2}$, problem \eqref{transformation3} remains difficult to solve. The following lemma can be used to further transform it into a simpler form.
\begin{lemma}\label{Lemma_FP_Must}
By further introducing an auxiliary variable $\xi_k$ and applying a quadratic transform, problem \eqref{transformation3} is reformulated into the equivalent form \eqref{transformation4} at the bottom of this page. The optimal $\xi_k$ is given by $\xi_k^{\star}=\frac{\mathbf{h}_k^{bT} \mathbf{w}_k}
{1+\sum_{i=1}^{K}\left\lvert\mathbf{h}_k^{bT} \mathbf{w}_i\right\rvert^2}$.
\end{lemma}

\begin{figure*}[hb] 
\centering 
\vspace{-18pt}
\begin{subequations}\label{transformation4}
\begin{align}
&\max_{\mathbf{w}_k,\mathbf{x}^p,\tau_k,\mu_k,\nu_k, \xi_k}~\sum_{k=1}^{K}\tau_k\left(\log(1+\mu_k)-\mu_k+(1+\mu_k)\left(2\Re\left\{{\xi}_k^{*}
\mathbf{h}_k^{bT} \mathbf{w}_k\right\}
-\left\lvert\xi_k\right\rvert^2\left(1+\sum_{i=1}^{K}\left\lvert\mathbf{h}_k^{bT} \mathbf{w}_i\right\rvert^2
\right)\right)\right.\nonumber\\
&\quad\quad\quad\quad\quad\quad\quad\quad\quad\quad\quad+\left.\log(1+\nu_k)-\nu_k+(1+\nu_k)\frac{G_{\Gamma}-\Gamma_k}{1+G_{\Gamma}}-\log_2\left(1+G_{\Gamma}\right)\right)\label{transformation4_Objective}\\
&{\rm{s.t.}}~\eqref{C1},\eqref{C2},\eqref{C3},\eqref{C4}.\nonumber
\end{align}
\end{subequations}
\vspace{-27pt}
\end{figure*}

\begin{IEEEproof}
Since problem \eqref{transformation4} is a convex problem with respect to $\xi_k$ for fixed $\{\mathbf{w}_k,\tau_k,\mu_k,\nu_k\}_{k=1}^{K}$ and $\mathbf{x}^p$, we apply the first-order optimality condition, i.e.,
\begin{align}
\frac{\partial}{\partial \xi_k}\eqref{transformation4_Objective}=
\mathbf{h}_k^{bT} \mathbf{w}_k-\xi_k
\left(1+\sum_{i=1}^{K}\left\lvert\mathbf{h}_k^{bT} \mathbf{w}_i\right\rvert^2
\right)=0.
\end{align}
The optimal $\xi_k$ are easily seen as follows:
\begin{align}
\xi_k^{\star}=\frac{\mathbf{h}_k^{bT} \mathbf{w}_k}
{1+\sum_{i=1}^{K}\left\lvert\mathbf{h}_k^{bT} \mathbf{w}_i\right\rvert^2}.
\end{align}
Substituting $\xi_k^{\star}$ back into \eqref{transformation4_Objective} recovers the objective function in \eqref{transformation3_Objective}, thereby verifying the equivalence of these two problems.
\end{IEEEproof}
At this point, the transformed problem \eqref{transformation4} is a joint optimization problem involving $\{\mathbf{w}_k,\tau_k,\mu_k,\nu_k,\xi_k\}_{k=1}^{K}$ and $\mathbf{x}^p$, the BCD algorithm can be used to solve it. The optimal $\{\tau_k^{\star}, \mu_k^{\star}, \nu_k^{\star}, \xi_k^{\star}\}_{k=1}^{K}$ have been obtained, we next discuss how to  get the solution of $\mathbf{w}_k$ and $\mathbf{x}^p$. 

\subsection{Lagrange Multiplier Method for Baseband Beamformer}
The marginal problem for $\mathbf{w}_k$ is given by \eqref{final_FP_problem} at the bottom of next page, whose objective function can be rewritten as follows:
\begin{figure*}[hb] 
\centering 
\hrule 
\begin{subequations}\label{final_FP_problem}
\begin{align}
\mathcal{P}_{2-1}: \min_{\mathbf{w}_k}~&\sum_{k=1}^{K}\tau_k\left((1+\mu_k)\left(\left\lvert\xi_k\right\rvert^2\sum_{i=1}^{K}\left\lvert\mathbf{h}_k^{bT} \mathbf{w}_i\right\rvert^2-
2\Re\left\{{\xi}_k^{*}\mathbf{h}_k^{bT} \mathbf{w}_k\right\}\right)+
\frac{\sum_{j=1}^{J}\left\lvert\mathbf{h}_j^{eT} \mathbf{w}_k\right\rvert^2}{1+G_{\Gamma}}(1+\nu_k)\right)\\
{\rm{s.t.}}~&\eqref{C1}.\nonumber
\end{align}
\end{subequations}
\hrule 
\end{figure*}
\begin{align}
&f(\{\mathbf{w}_{k}\}_{k=1}^{K})=\sum_{k=1}^{K}\Bigg(\sum_{i=1}^{K}\tau_{i}(1+\mu_{i})\left\lvert\xi_{i}\right\rvert^2\left\lvert\mathbf{h}_i^{bT} \mathbf{w}_k\right\rvert^2 \nonumber\\
&\!\!+\!\!\frac{1+\nu_k}{1+G_{\Gamma}}\sum_{j=1}^{J}\left\lvert\mathbf{h}_j^{eT} \mathbf{w}_k\right\rvert^2\!\!-\!\!2\tau_k(1+\mu_k)\Re\left\{{\xi}_k^{*}\mathbf{h}_k^{bT} \mathbf{w}_k\right\}\Bigg).
\end{align}
Since the objective function is convex and the constraint set is convex, the Karush-Kuhn-Tucker (KKT) conditions are sufficient for optimality. The Lagrangian function is given by
\begin{align}\label{Lagrange_Function}
{\mathcal{L}}=f(\{\mathbf{w}_{k}\}_{k=1}^{K})+\lambda\left(\sum_{k=1}^{K}\| \mathbf{w}_{k}\|^2- P_T\right),
\end{align}
where $\lambda\geq0$ is the Lagrange multiplier. The stationarity condition is derived by setting the complex derivative of \eqref{Lagrange_Function} to zero, which can be written as follows:
\begin{align}
&\frac{\partial{\mathcal{L}}}{\partial \mathbf{w}_{k}}=\frac{\partial{f}}{\partial \mathbf{w}_k}+\lambda \mathbf{w}_k=\sum_{i=1}^{K}\tau_{i}(1+\mu_{i})\left\lvert\xi_{i}\right\rvert^2 \mathbf{h}_i^{b*}\mathbf{h}_i^{bT} \mathbf{w}_k \\
&+\frac{1+\nu_k}{1+G_{\Gamma}}\sum_{j=1}^{J} \mathbf{h}_j^{e*}\mathbf{h}_j^{eT} \mathbf{w}_k -\tau_k(1+\mu_k){\xi}_k \mathbf{h}_k^{b*}+\lambda \mathbf{w}_{k}=0.\nonumber
\end{align}
It follows that
\begin{align}
&\left(\sum_{i=1}^{K}\tau_{i}(1+\mu_{i})\left\lvert\xi_{i}\right\rvert^2 \mathbf{h}_i^{b*}\mathbf{h}_i^{bT} + \frac{1+\nu_k}{1+G_{\Gamma}}\sum_{j=1}^{J} \mathbf{h}_j^{e*}\mathbf{h}_j^{eT} + \lambda \mathbf{I}_N\right)\nonumber\\
&\quad\cdot\mathbf{w}_k=\tau_k(1+\mu_k){\xi}_k \mathbf{h}_k^{b*},
\end{align}
which yields
\begin{equation}\label{w_k}
\begin{split}
\mathbf{w}_k = \tau_k(1+\mu_k){\xi}_k\Bigg(\sum_{i=1}^{K}\tau_{i}(1+\mu_{i})\left\lvert\xi_{i}\right\rvert^2 \mathbf{h}_i^{b*}\mathbf{h}_i^{bT} \\
+ \frac{1+\nu_k}{1+G_{\Gamma}}\sum_{j=1}^{J} \mathbf{h}_j^{e*}\mathbf{h}_j^{eT} + \lambda \mathbf{I}_N\Bigg)^{-1}\mathbf{h}_k^{b*}, \forall k.
\end{split}
\end{equation}

Then, it can be easily proved that $\|\mathbf{w}_{k}\|^2$ is a monotone decreasing function with respect to $\lambda$, and so is $\sum_{k=1}^{K}\| \mathbf{w}_{k}\|^2$. Therefore, we can use binary search to find $\lambda$, which is the solution to the constraint $\sum_{k=1}^{K}\| \mathbf{w}_{k}\|^2=P_T$.

\subsection{Gauss-Seidel Approach-Based Pinching Beamformer}
We now consider how to optimize the positions of PAs. Recalling \eqref{transformation4} and given $\{\mathbf{w}_k,\tau_k, \mu_k, \nu_k, \xi_k\}_{k=1}^{K}$, the marginal problem with respect to $\mathbf{x}^p$ is given by \eqref{position_problem}, which is shown at the bottom of next page.
\begin{figure*}[hb] 
\centering 
\vspace{-18pt}
\begin{subequations}\label{position_problem}
\begin{align}
\min_{\mathbf{x}^p}~&\sum_{k=1}^{K}\Bigg(\sum_{i=1}^{K}\tau_{i}(1+\mu_{i})\left\lvert\xi_{i}\right\rvert^2\left\lvert\mathbf{h}_i^{bT}(\mathbf{x}^p) \mathbf{w}_k\right\rvert^2
+\tau_{k}\log_2\left(1+G_{\Gamma}(\mathbf{x}^p)\right)-\tau_{k}(1+\nu_k)\frac{G_{\Gamma}(\mathbf{x}^p)-\sum_{j=1}^{J}\left\lvert\mathbf{h}_j^{eT}(\mathbf{x}^p) \mathbf{w}_k\right\rvert^2}{1+G_{\Gamma}(\mathbf{x}^p)}\nonumber\\
&\quad\quad\quad\quad-2\tau_k(1+\mu_k)\Re\left\{{\xi}_k^{*}\mathbf{h}_k^{bT}(\mathbf{x}^p) \mathbf{w}_k\right\}\Bigg)\label{position_problem_Objective}\\
&{\rm{s.t.}}~\eqref{C2},\eqref{C3}.\nonumber
\end{align}
\end{subequations}
\end{figure*}
It can be further written in a more compact form as follows:
\begin{subequations}\label{position_problem_transform}
\begin{align}
\min_{\mathbf{x}^p}~&\text{tr}\left(\mathbf{W}\mathbf{W}^H\mathbf{H}^{b*}\mathbf{B}\mathbf{H}^{bT}\right) -2\Re\left\{\text{tr}\left(\mathbf{W}\mathbf{A}^H\mathbf{H}^{bT}\right)\right\}\nonumber\\
&+\frac{\text{tr}\left(\mathbf{W}\mathbf{C}\mathbf{W}^H\mathbf{H}^{e*}\mathbf{H}^{eT}\right)-\text{tr}\left(\mathbf{C}\right)G_{\Gamma}(\mathbf{x}^p)}{1+G_{\Gamma}(\mathbf{x}^p)}\nonumber\\
&+\text{tr}\left(\mathbf{\Upsilon}\right)\log_2\left(1+G_{\Gamma}(\mathbf{x}^p)\right)\label{position_problem_Objective}\\
&{\rm{s.t.}}~\eqref{C2},\eqref{C3}.\nonumber
\end{align}
\end{subequations}
where $\mathbf{H}^b=\left[\mathbf{h}_1^b(\mathbf{x}^p), \mathbf{h}_2^b(\mathbf{x}^p), \ldots, \mathbf{h}_K^b(\mathbf{x}^p)\right]\in \mathbb{C}^{N\times K}$,  $\mathbf{H}^e=[\mathbf{h}_1^e(\mathbf{x}^p), \mathbf{h}_2^e(\mathbf{x}^p), \ldots, \mathbf{h}_J^e(\mathbf{x}^p)]\in \mathbb{C}^{N\times J}$, and $\mathbf{W}=[\mathbf{w}_1, \mathbf{w}_2, \ldots, \mathbf{w}_K]\in \mathbb{C}^{N\times K}$. Furthermore, we introduce the auxiliary matrices $\mathbf{A} = \mathbf{\Upsilon}\mathbf{\Psi}\mathbf{\Xi}$, $\mathbf{B} = \mathbf{\Xi}\mathbf{\Upsilon}\mathbf{\Psi}\mathbf{\Xi}^H$, and $\mathbf{C} = \mathbf{\Upsilon}\mathbf{\Lambda}$, where 
$\mathbf{\Upsilon}=\text{diag}\{\tau_1^{\star}, \tau_2^{\star}, \dots, \tau_K^{\star}\}$, $\mathbf{\Psi}=\text{diag}\{1+\mu_1^{\star}, 1+\mu_2^{\star}, \dots, 1+\mu_K^{\star}\}$, $\mathbf{\Xi}=\text{diag}\{\xi_1^{\star}, \xi_2^{\star}, \dots, \xi_K^{\star}\}$, and $\mathbf{\Lambda}=\text{diag}\{1+\nu_1^{\star}, 1+\nu_2^{\star}, \ldots,1+\nu_K^{\star}\}$. 

Since the positions $\{x_{m,n}^p, \forall m,n\}$ are mutually coupled, obtaining the global optimal solution is challenging.
A Gauss-Seidel approach can be adopted to tackle this problem \cite{ref122}. 
When optimizing $x_{m,n}^p$, the positions of other PAs, $x_{m',n'}^p,\forall m' \neq m, n' \neq n$, are kept fixed. The marginal problem with respect to the scalar $x_{m,n}^p$ can be fomulated as follows:
\begin{subequations}\label{position_problem_transform_xn}
\begin{align}
\min_{x_{m,n}^p}~&\text{tr}\left(\mathbf{E}\widetilde{\mathbf{H}}^{bH}(x_{m,n}^p)\mathbf{B}\widetilde{\mathbf{H}}^{b}(x_{m,n}^p)\right) -2\Re\left\{\text{tr}\left(\mathbf{F}\widetilde{\mathbf{H}}^{b}(x_{m,n}^p)\right)\right\}\nonumber\\
&+\frac{\text{tr}\left(\mathbf{J}\widetilde{\mathbf{H}}^{eH}(x_{m,n}^p)\widetilde{\mathbf{H}}^{e}(x_{m,n}^p)\right)-\text{tr}\left(\mathbf{C}\right)G_{\Gamma}(x_{m,n}^p)}{1+G_{\Gamma}(x_{m,n}^p)}\nonumber\\
&+\text{tr}\left(\mathbf{\Upsilon}\right)\log_2\left(1+G_{\Gamma}(x_{m,n}^p)\right)\label{position_problem_Objective}\\
&{\rm{s.t.}}~\eqref{C2},\eqref{C3}.\nonumber
\end{align}
\end{subequations}
where we define $\mathbf{E}=\mathbf{W}\mathbf{W}^H$, $\mathbf{F}=\mathbf{W}\mathbf{A}^H$, and $\mathbf{J}=\mathbf{W}\mathbf{C}\mathbf{W}^H$. Matrices $\widetilde{\mathbf{H}}^b$ and $\widetilde{\mathbf{H}}^e$ denote the transposes of matrices $\mathbf{H}^b$ and $\mathbf{H}^e$, respectively. Specifically, $\widetilde{\mathbf{H}}^b(x_{m,n}^p)\in \mathbb{C}^{K\times N}$ is defined as follows:
\begin{equation}
    \widetilde{\mathbf{H}}^b(x_{m,n}^p) \!\!= \!\!\left[\widetilde{\mathbf{h}}_1^b, \dots, \widetilde{\mathbf{h}}_{n-1}^b, \widetilde{\mathbf{h}}_n^b (x_{m,n}^p), \widetilde{\mathbf{h}}_{n+1}^b, \dots, \widetilde{\mathbf{h}}_N^b\right],
\end{equation}
with $\widetilde{\mathbf{h}}_n^b (x_{m,n}^p) \in \mathbb{C}^{K \times 1}$ representing the $n $-th column of $\mathbf{H}^{bT}$, whose $k$-th entry is given by
\begin{equation}\label{hn1}
[\widetilde{\mathbf{h}}_n^b (x_{m,n}^p)]_k = \Pi_k^b(x_{m,n}^p)+\sum_{m'\neq m}^{M_n}\Pi_k^b(x_{m',n}^p),
\end{equation}
where $\Pi_k^b(x_{m,n}^p)$ is denoted as 
\begin{equation}\label{hn}
\Pi_k^b(x_{m,n}^p) = \frac{\sqrt{\eta} e^{-j 2 \pi \left( \frac{\|\boldsymbol{\psi}^b_k - \boldsymbol{\psi}_{m,n}^p\|}{\lambda_c}+\frac{x_{m,n}^p-x_{0,n}^p}{\lambda_p}\right)}}{\sqrt{M_n}\|\boldsymbol{\psi}^b_k - \boldsymbol{\psi}_{m,n}^p\|}.
\end{equation}
Similarly, $\widetilde{\mathbf{H}}^e(x_{m,n}^p)\in \mathbb{C}^{J\times N}$ is defined as follows: 
\begin{equation}
\widetilde{\mathbf{H}}^e(x_{m,n}^p)\! \!= \!\!\left[\widetilde{\mathbf{h}}_1^e, \dots, \widetilde{\mathbf{h}}_{n-1}^e, \widetilde{\mathbf{h}}_n^e (x_{m,n}^p), \widetilde{\mathbf{h}}_{n+1}^e, \dots, \widetilde{\mathbf{h}}_N^e\right], 
\end{equation}
and the $j$th entry of $\widetilde{\mathbf{h}}_n^e (x_{m,n}^p)$ is given by
\begin{equation}\label{gn1}
[\widetilde{\mathbf{h}}_n^e (x_{m,n}^p)]_j = \Pi_j^e(x_{m,n}^p)+\sum_{m'\neq m}^{M_n}\Pi_j^e(x_{m',n}^p),
\end{equation}
where $\Pi_j^e(x_{m,n}^p)$ is represented as 
\begin{equation}\label{gn}
\Pi_j^e(x_{m,n}^p) =\frac{\sqrt{\eta} e^{-j2 \pi \left( \frac{\|\boldsymbol{\psi}^e_j - \boldsymbol{\psi}_{m,n}^p\|}{\lambda_c}+\frac{x_{m,n}^p-x_{0,n}^p}{\lambda_p}\right)}}{\sqrt{M_n}\|\boldsymbol{\psi}^e_j - \boldsymbol{\psi}_{m,n}^p\|}.
\end{equation}
The matrix $G_{\Gamma}(x_{m,n}^p)$ is then defined as 
\begin{align}
G_{\Gamma}(x_{m,n}^p)=P_T\Bigg(\|\widetilde{\mathbf{H}}_{/n}^{e}\|_{\mathrm{F}}^2+\|\widetilde{\mathbf{h}}_n^e (x_{m,n}^p)\|^2\Bigg),
\end{align}
where $\widetilde{\mathbf{H}}_{/n}^{e}$ denotes $\widetilde{\mathbf{H}}^{e}$ with the $n$th column removed.

By simple lines of derivation, the problem in \eqref{position_problem_transform_xn} can be rewritten as \eqref{position_problem_transform_xn1} at the bottom of next page,
\begin{figure*}[hb]
\centering 
\hrule 
\begin{subequations}\label{position_problem_transform_xn1} 
\begin{align}
\min_{x_{m,n}^p}~&[\mathbf{E}]_{n,n}\widetilde{\mathbf{h}}_n^{bH}(x_{m,n}^p)\mathbf{B}\widetilde{\mathbf{h}}_n^{b}(x_{m,n}^p)
+2\Re\left\{\mathbf{a}_n^T\widetilde{\mathbf{h}}_n^{b}(x_{m,n}^p)\right\} +\text{tr}\left(\mathbf{\Upsilon}\right)\log_2\left(P_T\|\widetilde{\mathbf{h}}_n^e (x_{m,n}^p)\|^2+C_1\right)\nonumber\\
&+\frac{\left([\mathbf{J}]_{n,n}-P_T\text{tr}\left(\mathbf{C}\right)\right)\|\widetilde{\mathbf{h}}_n^{e}(x_{m,n}^p)\|^2+2\Re\left\{\mathbf{b}_n^T\widetilde{\mathbf{h}}_n^{e}(x_{m,n}^p)\right\}+C_2}{P_T\|\widetilde{\mathbf{h}}_n^e (x_{m,n}^p)\|^2+C_1}\label{position_problem_Objective}\\
&{\rm{s.t.}}~\eqref{C2},\eqref{C3}.\nonumber
\end{align}
\end{subequations}
\vspace{-25pt}
\end{figure*}
where $\mathbf{a}_n \in \mathbb{C}^{K\times 1}$ and $\mathbf{b}_n \in \mathbb{C}^{J\times 1}$ are defined as follows: 
\begin{align}
\mathbf{a}_n =\sum_{n'\neq n}[\mathbf{E}]_{n,n'}\mathbf{B}^T\widetilde{\mathbf{h}}_{n'}^{b*}-\mathbf{f}_n,
\end{align}
\begin{align}
\mathbf{b}_n =\sum_{n'\neq n}[\mathbf{J}]_{n,n'}\widetilde{\mathbf{h}}_{n'}^{e*},
\end{align}
with $\mathbf{f}_n^T \in \mathbb{C}^{1\times K}$ representing the $n$-th row of $\mathbf{F}$. The constant terms $C_1$ and $C_2$ are given as follows:
\begin{align}
C_1 = 1+ P_T\|\widetilde{\mathbf{H}}_{/n}^{e}\|_{\mathrm{F}}^2,
\end{align}
\begin{align}
C_2 = \text{tr}\left(\bar{\mathbf{J}} \widetilde{\mathbf{H}}_{/n}^{eH} \widetilde{\mathbf{H}}_{/n}^{e} \right) -P_T\text{tr}\left(\mathbf{C}\right)\|\widetilde{\mathbf{H}}_{/n}^{e}\|_{\mathrm{F}}^2,
\end{align}
with matrix $\bar{\mathbf{J}}$ denoting the submatrix of $\mathbf{J}$ obtained by deleting its $n$th row and $n$-th column.

By inspecting \eqref{position_problem_Objective}, it can be observed that the PA position variable $x_{m,n}^p$ only appears in $\widetilde{\mathbf{h}}_{n}^{b}(x_{m,n}^p)$ and $\widetilde{\mathbf{h}}_{n}^{e}(x_{m,n}^p)$. However, each also contains components that are unrelated to the $m$th PA. To further simplify the formulation, we can decompose them as follows:
\begin{subequations}\label{position_decompose}
\begin{align}
\widetilde{\mathbf{h}}_{n}^{b}(x_{m,n}^p) = \mathbf{\Pi}^b + \mathbf{\Pi}_{\mathrm{const}}^b,\\
\widetilde{\mathbf{h}}_{n}^{e}(x_{m,n}^p) = \mathbf{\Pi}^e + \mathbf{\Pi}_{\mathrm{const}}^e,
\end{align}
\end{subequations}
where 
\begin{subequations}\label{position_decompose_1}
\begin{align}
&\mathbf{\Pi}^b=\left[\Pi_1^b(x_{m,n}^p), \Pi_2^b(x_{m,n}^p),\ldots, \Pi_K^b(x_{m,n}^p) \right]^T,\\
&\mathbf{\Pi}_{\mathrm{const}}^b \!\!=\!\!\left[\sum_{m'\neq m}^{M_n}\Pi_1^b(x_{m',n}^p), \dots, \sum_{m'\neq m}^{M_n}\Pi_K^b(x_{m',n}^p)\right]^T,\\
&\mathbf{\Pi}^e=\left[\Pi_1^e(x_{m,n}^p), \Pi_2^e(x_{m,n}^p),\ldots, \Pi_J^e(x_{m,n}^p) \right]^T,\\
&\mathbf{\Pi}_{\mathrm{const}}^e\!\!=\!\!\left[\sum_{m'\neq m}^{M_n}\Pi_1^e(x_{m',n}^p), \dots, \sum_{m'\neq m}^{M_n}\Pi_J^e(x_{m',n}^p)\right]^T.
\end{align}
\end{subequations}

Based on the above definitions, \eqref{position_problem_Objective} can be further reformulated as \eqref{position_problem_Objective_2}, which is presented at the bottom of this page,
\begin{figure*}[hb]
\centering 
\hrule 
\begin{align}
f(x_{m,n}^p)=~&[\mathbf{E}]_{n,n}\mathbf{\Pi}^{bH}\mathbf{B}\mathbf{\Pi}^b
+2\Re\left\{\bar{\mathbf{a}}_n^T\mathbf{\Pi}^b\right\} +\text{tr}\left(\mathbf{\Upsilon}\right)\log_2\left(P_T\left(\|\mathbf{\Pi}^{e}\|^2+2\Re \left\{\mathbf{\Pi}_{\mathrm{const}}^{eH}\mathbf{\Pi}^e\right\}\right)+\bar{C}_1\right)\nonumber\\
&+\frac{\left([\mathbf{J}]_{n,n}-P_T\text{tr}\left(\mathbf{C}\right)\right)\|\mathbf{\Pi}^e\|^2+2\Re\left\{\bar{\mathbf{b}}_n^T\mathbf{\Pi}^e\right\} +\bar{C}_2}{P_T\left(\|\mathbf{\Pi}^{e}\|^2+2\Re \left\{\mathbf{\Pi}_{\mathrm{const}}^{eH}\mathbf{\Pi}^e\right\}\right)+\bar{C}_1}.\label{position_problem_Objective_2}
\end{align}
\vspace{-20pt}
\end{figure*}
where the newly introduced variables $\bar{\mathbf{a}}_n$, $\bar{\mathbf{b}}_n$, $\bar{C}_1$ and $\bar{C}_2$ are given by:
\begin{subequations}
\begin{align}
&\bar{\mathbf{a}}_n = \mathbf{a}_n + [\mathbf{E}]_{n,n}\mathbf{B}^T\mathbf{\Pi}_{\mathrm{const}}^{b*},\\
&\bar{\mathbf{b}}_n = \mathbf{b}_n + ([\mathbf{J}]_{n,n}-P_T\text{tr}\left(\mathbf{C}\right))\mathbf{\Pi}_{\mathrm{const}}^{e*},\\
&\bar{C}_1 = C_1 + P_T\|\mathbf{\Pi}_{\mathrm{const}}^{e}\|^2,\\
&\bar{C}_2 = C_2 + ([\mathbf{J}]_{n,n}-P_T\text{tr}\left(\mathbf{C}\right))\|\mathbf{\Pi}_{\mathrm{const}}^{e}\|^2+2\Re\left\{\mathbf{b}_n^T\mathbf{\Pi}_{\mathrm{const}}^{e}\right\}. 
\end{align}
\end{subequations}
Therefore, the optimization problem for the position of the $m$th PA on the $n$th waveguide can be formulated as follows:
\begin{subequations}\label{fxmnp}
\begin{align} \label{xp}
\mathcal{P}_{2-2}: &\min_{x_{m,n}^p} f (x_{m,n}^p) \\
&{\rm{s.t.}}~\eqref{C2},\eqref{C3}.\nonumber
\end{align}
\end{subequations}
Observing the objective function in \eqref{xp}, it is significantly more complex compared to the objective function in \eqref{problem_for_T1}, as it contains numerous exponential summations, leading to a substantial increase in the number of stationary points. The gradient-based method is no longer effective in this case \cite{ref10}, whereas a one-dimensional search method proves to be a viable approach for finding a locally optimal solution. 
By discretizing the interval $[-D/2, D/2]$ into $N_s$ sample points, with a segment length of $\Delta = \frac{D}{N_s-1}$, we define the set of candidate positions as,
\begin{equation}
    \mathcal{X} \triangleq \left\{ -\frac{D}{2} + i\Delta \;\middle|\; i = 0,1,\dots,N_s - 1 \right\}.
\end{equation}
An approximate optimal position $x_{m,n}^p$ can be obtained by selecting  
\begin{equation}\label{position}
x_{m,n}^p= \arg\min_{x_{m,n}^p \in \mathcal{X}/\mathcal{X}(\mathcal{I})} f (x_{m,n}^p),
\end{equation}
where $/$ denotes the set difference, and $\mathcal{X}(\mathcal{I})$ represents the set $\mathcal{X}$ corresponding to the indices $i \in \mathcal{I}$. The set $\mathcal{I}$ is defined as follows:
\begin{equation}
\mathcal{I} =
\begin{cases}
  \displaystyle
  \bigcup_{m'=1}^{m-1} 
  \left\{ i \;\middle|\;
  i \in \left\{ i_{m',n}^{\rm floor}, 
  \dots, 
  i_{m',n}^{\rm ceil} \right\} 
  \right\}, & m \neq 1 \\
  \varnothing, & m=1,
\end{cases}
\end{equation}
where $i_{m',n}^{\rm floor}\!\!=\!\!\left\lfloor \frac{2x_{m',n}^p+D-2\Delta_{\rm min}}{2\Delta} \right\rfloor$, $i_{m',n}^{\rm ceil}\!\!=\!\!\left\lceil \frac{2x_{m',n}^p+D+2\Delta_{\rm min}}{2\Delta} \right\rceil$, ensuring that the distance between each PA on the same waveguide satisfies the constraint \eqref{C3}. By using \eqref{position} to sequentially update $\{x_{m,n}^p, \forall m,n\}$, a locally optimal solution to \eqref{fxmnp} can be obtained. 

\begin{algorithm}[t!]
\caption{FP-BCD Algorithm for WSSR Maximization Problem $\mathcal{P}_{1}$ in PASS-based Communications}\label{Alg2}
\begin{algorithmic}[1]
    \State Initialize $(\mathbf{x}^p)^0 = [(\mathbf{x}_1^p)^0, \ldots, (\mathbf{x}_N^p)^0]^T$, the iteration count $t=0$, the maximum iteration number $T$. 
    \State Initialize $\mathbf{w}_k$ by using the MRT algorithm, i.e., $\mathbf{w}_k = \frac{\sqrt{P_T}\mathbf{h}_k^{b*}}{\|\mathbf{h}_k^b\|}$, and $\tau_k^0$ can be obtained.
    \Repeat
        \State Update $(\mu_k)^{t}$ by $\mu_k=\frac{\left\lvert \mathbf{h}_k^{bT} \mathbf{w}_k\right\rvert^2}{1+\sum_{i=1,i\ne k}^{K}\left\lvert\mathbf{h}_k^{bT} \mathbf{w}_i\right\rvert^2}$;
        \State Update $(\nu_k)^{t}$ by $\nu_k=\frac{G_{\Gamma}-\Gamma_k}{1+\Gamma_k}$;
        \State Update $(\xi_k)^{t}$ by $\xi_k=\frac{\mathbf{h}_k^{bT} \mathbf{w}_k}{1+\sum_{i=1}^{K}\left\lvert\mathbf{h}_k^{bT} \mathbf{w}_i\right\rvert^2}$;
        \State Update $(\mathbf{w}_k)^t$ by \eqref{w_k}, and $\lambda$ is obtained by binary search.
        \State Update $(\mathbf{x}^p)^t$ by sequentially executing \eqref{xp} $M$ times.
        \State Update $(\tau_k)^{t}$ by \eqref{bk} and calculate the WSSR $(R^{\mathrm{sec}})^t$.
    \Until{the increment of $R^{\mathrm{sec}}$ becomes smaller than a predefined threshold or reaches the maximum iteration count $T$.}
\end{algorithmic}
\end{algorithm}
The overall FP-BCD algorithm for solving the WSSR maximization problem is summarized in \textbf{Algorithm \ref{Alg2}}, which is guaranteed to converge to a stationary point solution \cite{ref121}. The main computational complexity arises from the updates of variables $\mu_k$, $\nu_k$, $\xi_k$, $\mathbf{w}_k$, and $\mathbf{x}^p$. Specifically, the complexities of updating $\mu_k$, $\nu_k$, $\xi_k$, and $\mathbf{x}^p$ are in order of $\mathcal{O}(K^2N)$, $\mathcal{O}(KJN)$, $\mathcal{O}(K^2N)$, and $\mathcal{O}(N_sM(K+J))$, respectively, primarily due to vector multiplications. The update of $\mathbf{w}_k$ involves matrix inversion and bisection search, contributing a complexity of $\mathcal{O}(K(N^3 + N^2 + \log(1/\epsilon)))$, where $\epsilon$ denotes the bisection accuracy.

\section{Numerical Results}
In this section, numerical results are presented to evaluate the advantages of PASS and verify the effectiveness of the proposed algorithms. The number of Monte-Carlo simulations is set to 500. Unless stated otherwise, the following simulation parameters are used: 
The height of all waveguides is set to $d = 3 \, \text{m}$, with their length matching the side length $D$, ensuring complete area coverage. The $y$-axis coordinates of the $n$th waveguide is set to $y_n^p=-\frac{D}{2}+\frac{nD}{N}$. To prevent mutual coupling, the spacing between PAs on the same waveguide is set to $\Delta_{\rm min} = \frac{\lambda_c}{2}$. The noise power at the receiver is $\sigma_{b,k}^2 = \sigma_{e,j}^2 = -90$ dBm. The carrier frequency is set to $f_c = 28$ GHz, and the waveguide wavelength is $\lambda_p = \frac{\lambda_c}{n_{\text{eff}}}$, where $n_{\text{eff}} = 1.4$ represents the effective refractive index of the dielectric waveguide.
For the proposed \textbf{Algorithm \ref{alg1}}, the initial step size is set to $\beta_{\text{ini}} = 10$\footnote{The initial step size is empirically determined through computer simulations. Different values may lead to convergence to different local optima, but the performance remains comparable to that of exhaustive search, with significantly reduced runtime.}, and the minimum tolerance step size is $\beta_{\text{min}} = 1 \times 10^{-13}$. The initial $x$-axis coordinates of the PAs on all waveguides are $\mathbf{x}^p = [0, \ldots, 0]\in \mathbb{R}^N$. 
For \textbf{Algorithm \ref{Alg2}}, when each waveguide activates a single PA, the $x$-axis coordinate is initialized to $\mathbf{x}^p = [0, \ldots, 0]\in \mathbb{R}^N$. In cases where multiple PAs are activated per waveguide, the $x$-axis coordinates are randomly initialized and are referred to as ``Pinching Antenna mul.'' in the following figures. Furthermore, the baseband beamforming vector $\mathbf{w}_k$ is initialized by MRT algorithm.
For performance comparison, a traditional fixed-antenna (FA) BS is adopted. It is equipped with $N$ antennas, which are arranged along the $x$-axis with a half-wavelength spacing. The central position is set at $(0, 0, d)$.

\begin{figure}[!t]
\vspace{-10pt}
    \centering
    \includegraphics[width=0.35\textwidth]{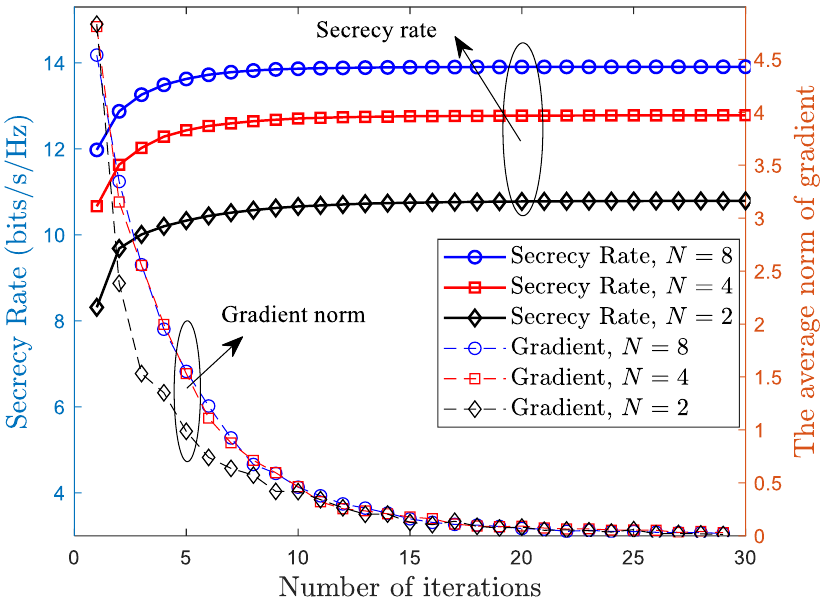} 
    \caption{The convergence curve of \textbf{Algorithm 1}, $P_T=20\,\text{dBm}$, $D = 30\, \text{m}$.}
    \label{fig:conver} 
\vspace{-10pt}
\end{figure}

\begin{figure}[t!]
\vspace{-10pt}
    \centering
    \subfigure[$D = 30\, \text{m}$.]{
        \includegraphics[width=0.22\textwidth]{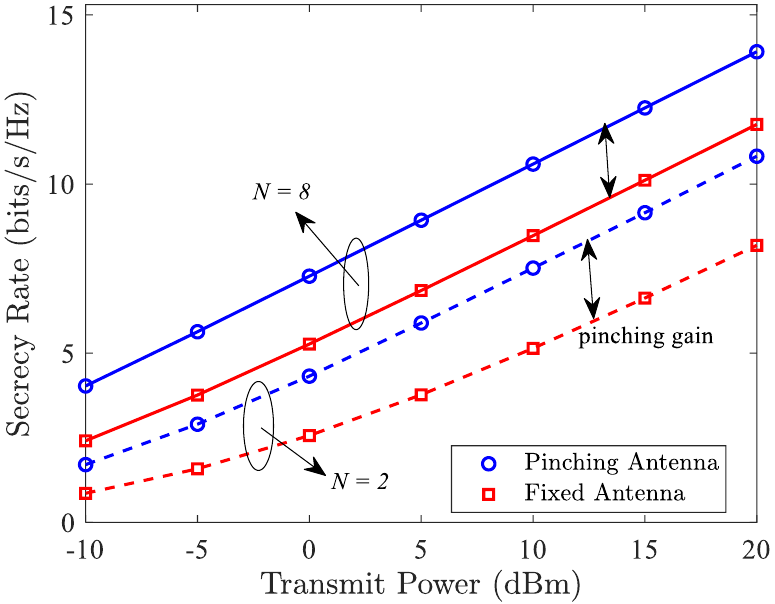}
        \label{fig:power}
    }
    \hfill
    \subfigure[$P_T=20\,\text{dBm}$.]{
        \includegraphics[width=0.22\textwidth]{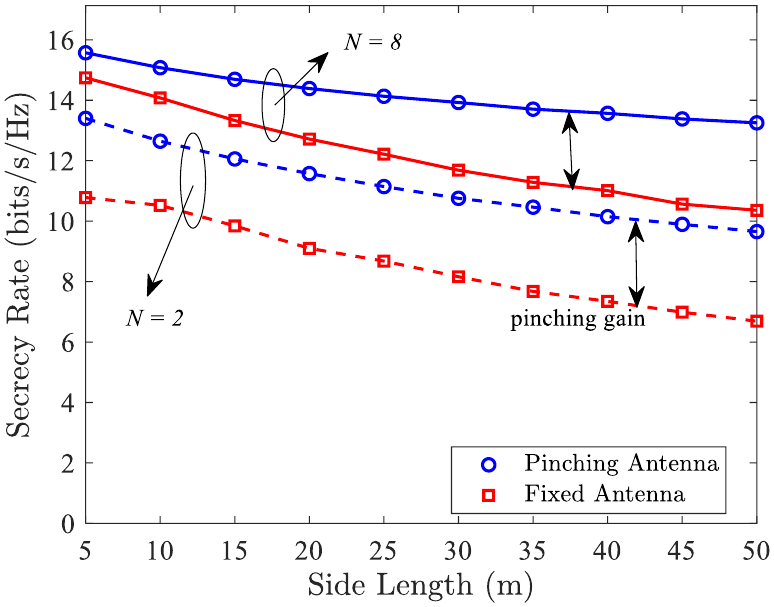}
        \label{fig:length}
    }
    \caption{The secrecy rate vs. transmit power and side length.}
    \label{fig:SR_single}
    \vspace{-8pt}
\end{figure}

\subsection{Secrecy Performance of Single-User PASS}
In this subsection, we present the performance of PASS in a scenario with a single Bob and a single Eve. 
Fig. \ref{fig:conver} demonstrates the convergence behavior of the gradient-based algorithm through the system secrecy rate and the average gradient norm $\nabla_{\rm norm}$. As shown in Fig. \ref{fig:conver}, the gradient-based method consistently exhibits a monotonic increase in the system secrecy rate under different numbers of waveguides. Furthermore, the convergence behavior of the algorithm can be validated through $\nabla_{\rm norm}$. After 20 iterations, the average gradient norm nearly drops to zero, further demonstrating the efficiency of the algorithm.

Fig. \ref{fig:power} shows the impact of transmit power on secrecy rate. Both PASS and conventional FA systems experience an increase in secrecy rate with higher $P_T$. Under all conditions, PA outperforms FA significantly due to strong LoS links, which mitigate large-scale path loss. This performance gain increases as transmit power rises. 
Fig. \ref{fig:length} further examines how secrecy rate varies with the side length of the user distribution area. 
As the side length increases, the secrecy rate gradually decreases, as the average distance between Bob and PA or FA increases, resulting in higher large-scale path loss. However, the performance gain of PA relative to FA increases with the side length, as PA can adjust its position flexibly, effectively reducing the impact of large-scale path loss.

\begin{figure}[!t]
\vspace{-10pt}
    \centering
    \includegraphics[width=0.33\textwidth]{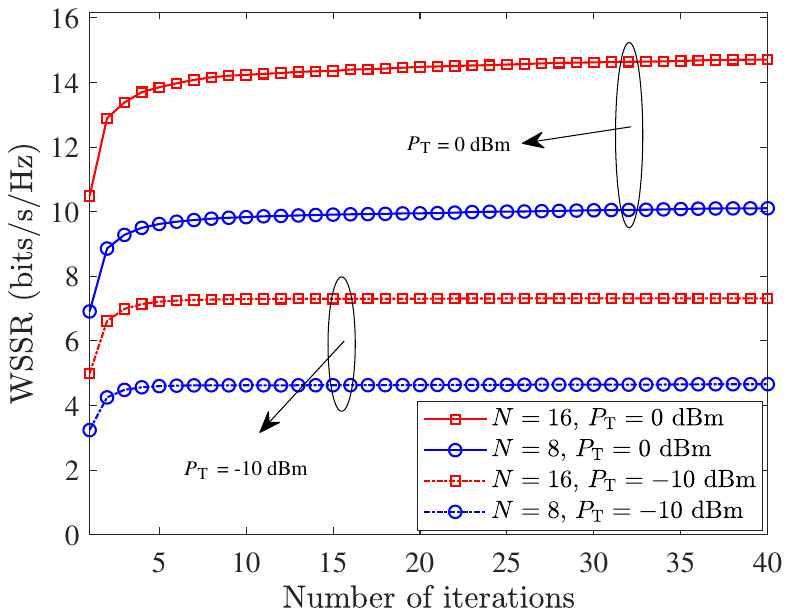} 
    \caption{The convergence curve of \textbf{Algorithm 2}, $D = 60\, \text{m}$, $N_s = 10^{4}$.}
    \label{fig:conver2} 
\vspace{-10pt}
\end{figure}
\begin{figure}[!t]
    \centering
    \includegraphics[width=0.33\textwidth]{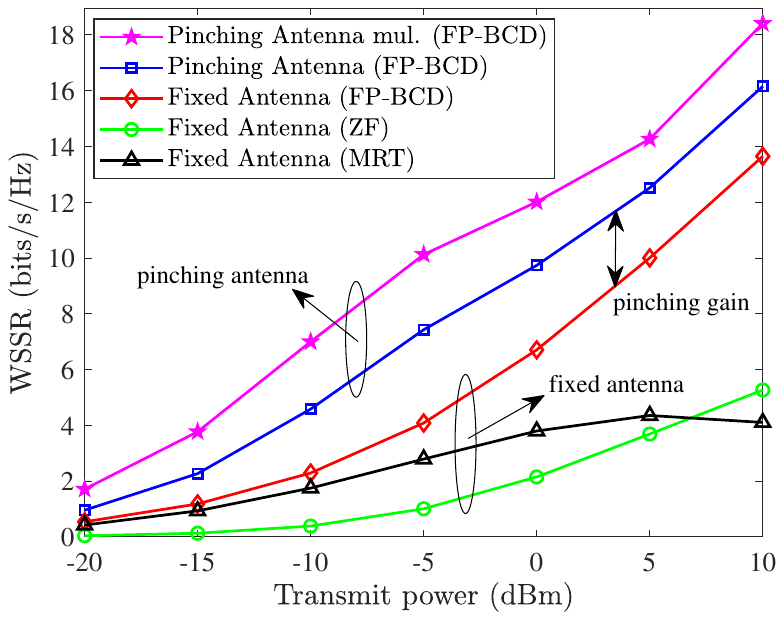} 
    \caption{The WSSR vs. transmit power, $D = 60\, \text{m}$, $N_s = 10^{4}$, $M_n =2$.}
    \label{fig:power2} 
\vspace{-10pt}
\end{figure}
\begin{figure}[!t]
\vspace{-10pt}
    \centering
    \includegraphics[width=0.33\textwidth]{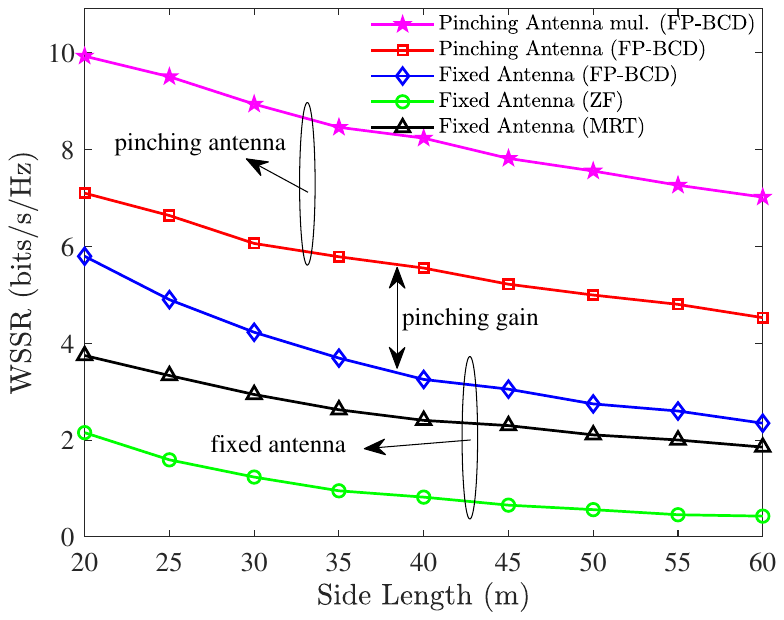} 
    \caption{The WSSR vs. side length, $P_T=-10\,\text{dBm}$, $N_s = 10^{4}$, $M_n =2$.}
    \label{fig:length2} 
\vspace{-10pt}
\end{figure}

\begin{figure}[t!]
\vspace{-5pt}
    \centering
    \subfigure[]{
        \includegraphics[width=0.22\textwidth]{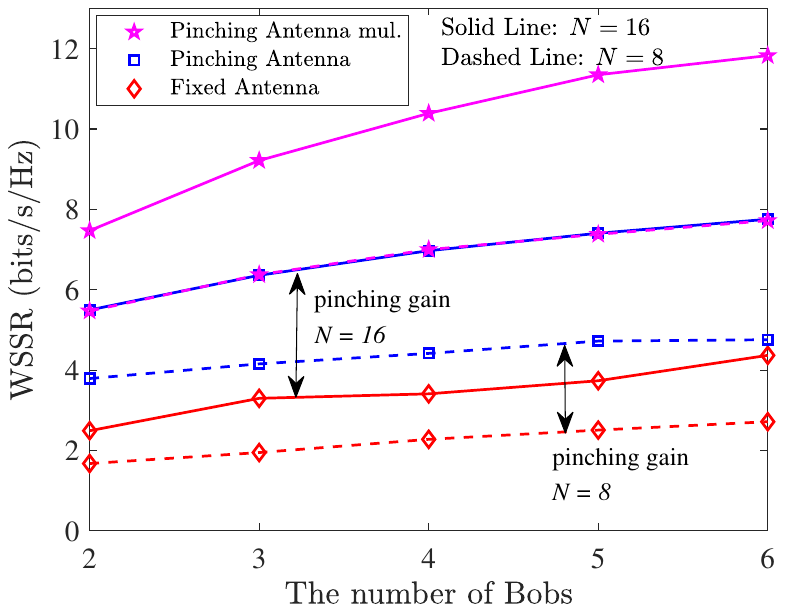}
        \label{fig:Bob}
    }
    \hfill
    \subfigure[]{
        \includegraphics[width=0.22\textwidth]{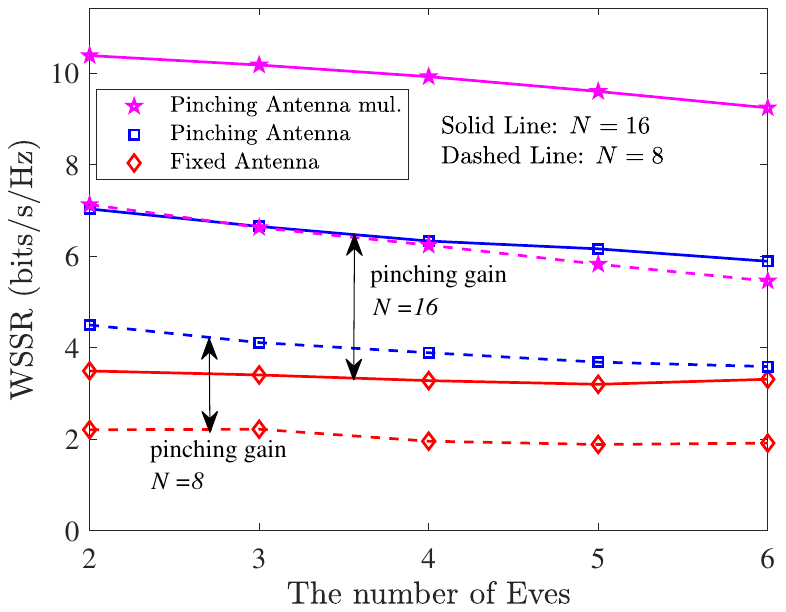}
        \label{fig:Eve}
    }
    \caption{The WSSR vs. the number of Bobs and Eves, $P_T=-10\,\text{dBm}$, $D = 60\, \text{m}$, $N_s = 10^{4}$, $M_n =2$.}
    \vspace{-8pt}
\end{figure}

\subsection{Secrecy Performance of Multiuser PASS}
In this subsection, we consider a scenario with multiple Bobs and multiple Eves. Unless otherwise specified, the number of waveguides is set to $N = 8$, the number of Bobs to $K = 4$, and the number of Eves to $J = 2$. 
In Fig. \ref{fig:conver2}, the convergence curve of the proposed FP-based alternating optimization algorithm is presented. Under different transmit power levels and varying numbers of waveguides, the WSSR increases monotonically with the number of iterations. Notably, the algorithm achieves stable convergence within 10 iterations which demonstrates its effectiveness.

Fig. \ref{fig:power2} illustrates the WSSR for various transmit power levels. To evaluate the effectiveness of the proposed FP-BCD algorithm, it is compared with classical linear beamforming schemes, namely MRT and ZF, in a FA system. The results show that the proposed algorithm outperforms the baseline schemes at all transmit power levels. Additionally, MRT performs better than ZF at low transmit power but becomes less effective as power increases due to the shift from a noise-dominated to an interference-dominated regime, where MRT fails to suppress interference. Furthermore, the PASS significantly improves the WSSR compared to the FA system. This improvement is attributed to the adjustment of the PA positions, which both reduces the large-scale fading of the legitimate channels and aligns the signal in the orthogonal space of the wiretap channel through phase control. Moreover, it is evident that increasing the number of PAs on each waveguide enhances the DoFs for pinching beamforming, further improving system performance.

\begin{figure}[!t]
\vspace{-10pt}
    \centering
    \includegraphics[width=0.33\textwidth]{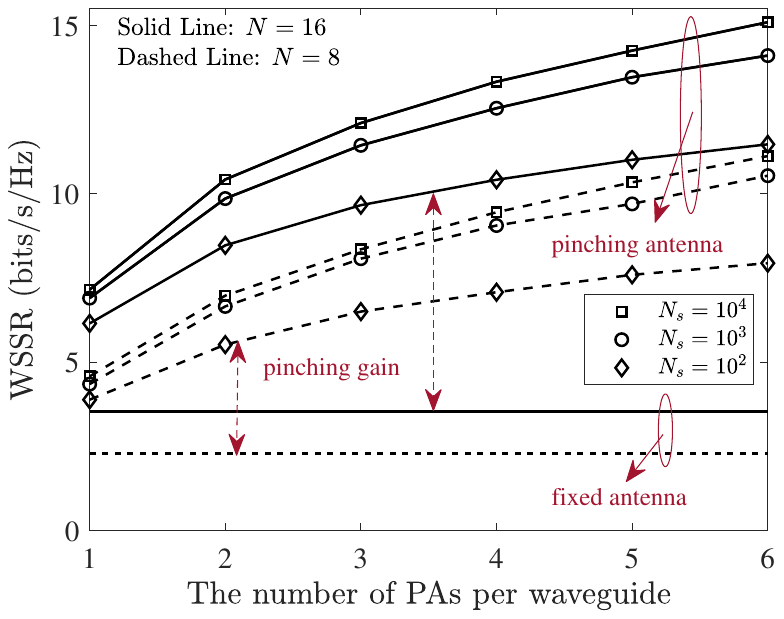} 
    \caption{The WSSR vs. the number of PAs per waveguide, $P_T=-10\,\text{dBm}$, $D = 60\, \text{m}$.}
    \label{fig:M} 
\vspace{-10pt}
\end{figure}
Fig. \ref{fig:length2} analyzes the impact of side length on system WSSR. Similar to the single-user case, both the PASS and FA systems experience a decrease in WSSR as the side length increases. This is due to the increased average distance between all Bobs and the BS transmit antennas, which leads to greater path loss, reducing the received signal strength. Also, PASS mitigates large-scale path loss by adjusting PAs' positions, leading to a slower decline in WSSR. Likewise, activating multiple PAs on each waveguide enhances the system's WSSR.

Fig. \ref{fig:Bob} and Fig. \ref{fig:Eve} further compares the effect of varying the number of Bobs and Eves on the system WSSR. In Fig. \ref{fig:Bob}, as the number of Bobs increases, the WSSR for both the PASS and FA systems gradually improves. Notably, the PASS system significantly outperforms the FA system, and the WSSR growth trend becomes steeper when multiple PAs are activated per waveguide. This can be attributed to two factors: first, the adjustment of PAs' positions mitigates large-scale path loss; second, the pinching beamformer effectively adjusts the signal phase to better suppress inter-Bob interference.
In Fig. \ref{fig:Eve}, the PASS consistently demonstrates a significant advantage at all Eve counts. By increasing the number of waveguides (antennas) or activating additional PAs, the WSSR can be further enhanced. However, as the number of Eves increases, the WSSR gradually declines in both systems. This is expected, as more Eves lead to an increased number of wiretap channels. Notably, the decline in WSSR is more pronounced in the PASS. This is because the increased number of Eves has a more significant impact on the pinching beamformer, where PA position adjustments directly influence the large-scale fading coefficient, further amplifying the downward trend of WSSR.

In Fig. \ref{fig:M}, we further illustrate the impact of the number of PAs per waveguide $M_n$ on WSSR and compare it with the FA system. It can be observed that the PASS significantly outperforms the FA system, and the performance gap between them widens as the number of PAs per waveguide $M_n$ increases. This is because a larger number of PAs facilitates enhanced cooperation which allows for more flexible signal phase adjustments to better align with the channels. Consequently, this improves interference suppression among Bobs and effectively reduces signal leakage to Eves.
Furthermore, increasing the sampling accuracy of the one-dimentional search to refine PAs' positions can further improve WSSR. However, this also leads to higher computational complexity. Therefore, in practical applications, an appropriate accuracy level should be selected based on the trade-off between computational cost and performance improvement.

\section{Conclusion}
This paper investigated the joint design of baseband beamforming and pinching beamforming for a PASS-enabled multiuser wiretap channel. In the single-user scenario, we derived a closed-form solution for the optimal baseband beamformer and proposed a gradient-based approach to optimize the positions of the PAs to maximize the secrecy rate. For the multiuser case, we proposed an FP-BCD algorithm leveraging the Gauss-Seidel approach combined with a one-dimensional search to design the joint beamforming. Numerical results verified that the proposed algorithms effectively optimize both the baseband beamforming and the pinching beamforming. Furthermore, PASS can utilize the additional DoFs provided by the flexible activation of PAs to significantly improve secrecy performance compared to conventional fixed-antenna systems. The proposed methods serve as promising candidates for further research into applying PASS to enhance PLS.

\appendices
\section{Proof of \textbf{Lemma~\ref{Lemma_for_mu_Delta}}}\label{Appendix_A}
The eigenvalues of $\mathbf{\Delta}$ can be obtained from the characteristic equation $\det(\mu \mathbf{I}-\mathbf{\Delta}) = 0$. 
Based on the fact that $(\mathbf{I}+\hat{\gamma}_e\mathbf{h}^{e*}\mathbf{h}^{eT})$ is a positive definite matrix, the following equivalence relationships can be obtained:
\begin{align}
\det(\mathbf{\Delta} - \mu \mathbf{I}) = 0 \iff 
\det((\mathbf{I}+\hat{\gamma}_e\mathbf{h}^{e*}\mathbf{h}^{eT})^{1/2})\nonumber \\
\cdot\det(\mu \mathbf{I}-\mathbf{\Delta}) \det((\mathbf{I}+\hat{\gamma}_e\mathbf{h}^{e*}\mathbf{h}^{eT})^{1/2}) = 0,
\end{align}
which yields $\det((\mu - 1) \mathbf{I} + \mu \hat{\gamma}_e \mathbf{h}^{e*}\mathbf{h}^{eT} - \hat{\gamma}_b\mathbf{h}^{b*}\mathbf{h}^{bT}) = 0.$
For simplicity, by defining $\mathbf{\Psi} \triangleq (\mu - 1) \mathbf{I} + \mu \hat{\gamma}_e \mathbf{h}^{e*}\mathbf{h}^{eT}$, we have
\begin{align}\label{characteristic equation}
\det(\mathbf{\Psi} - \hat{\gamma}_b \mathbf{h}^{b*}\mathbf{h}^{bT}) = 0.
\end{align}

If $\mu = 1$ is an eigenvalue of matrix $\mathbf{\Delta}$, it implies that the secrecy rate $R^{\rm sec} = 0$, which is not desirable. On the other hand, if $\mu \neq 1$, we can proceed with the following calculations.

Utilizing the matrix determinant lemma \cite{ref1}, we can reformulate \eqref{characteristic equation} as follows:
\begin{align}\label{characteristic_equation_1}
\det(\mathbf{\Psi})(1 - \hat{\gamma}_b \mathbf{h}^{bT} \mathbf{\Psi}^{-1} \mathbf{h}^{b*})  = 0,
\end{align}
where $\mathbf{\Psi}^{-1}$ is calculated using the Woodbury matrix identity \cite{ref1}, resulting in
\begin{align}\label{Psi_-1}
\mathbf{\Psi}^{-1} = \frac{1}{\mu - 1} \mathbf{I} - \frac{\mu \hat{\gamma}_e \mathbf{h}^{e*} \mathbf{h}^{eT}}{\left(\mu - 1 + \mu \hat{\gamma}_e \|\mathbf{h}^e\|^2\right)(\mu - 1)}. 
\end{align}

Applying the matrix determinant lemma again, we obtain
\begin{align}\label{det_Psi}
\det(\mathbf{\Psi}) = (\mu - 1)^{N-1} (\mu - 1 + \mu \hat{\gamma}_e \|\mathbf{h}^e\|^2). 
\end{align}

Substituting \eqref{Psi_-1} and \eqref{det_Psi} into \eqref{characteristic_equation_1} yields
\begin{align}
(\mu - 1&)^{N-2}((\mu - 1)^2+(\mu - 1)\mu\hat{\gamma}_e \|\mathbf{h}^e\|^2 - (\mu - 1)\hat{\gamma}_b \|\mathbf{h}^b\|^2 \nonumber\\
&-\mu \hat{\gamma}_b\hat{\gamma}_e \|\mathbf{h}^b\|^2\|\mathbf{h}^e\|^2 + \mu \hat{\gamma}_b\hat{\gamma}_e |\mathbf{h}^{bT}\mathbf{h}^{e*}|^2)=0,
\end{align}
which can be further expressed as follows:
\begin{align}\label{root_of_eigenvalue}
(\mu - 1)&^{N-2} \bigg[ (1+\hat{\gamma}_e \|\mathbf{h}^e\|^2)(\mu - 1)^2  
+ \Big( \hat{\gamma}_e \|\mathbf{h}^e\|^2 - \hat{\gamma}_b \|\mathbf{h}^b\|^2  \nonumber \\
&- \hat{\gamma}_b\hat{\gamma}_e \|\mathbf{h}^b\|^2\|\mathbf{h}^e\|^2  
+ \hat{\gamma}_b\hat{\gamma}_e |\mathbf{h}^{bT}\mathbf{h}^{e*}|^2 \Big) (\mu - 1)\nonumber\\
&- \hat{\gamma}_b\hat{\gamma}_e \|\mathbf{h}^b\|^2\|\mathbf{h}^e\|^2  
+ \hat{\gamma}_b\hat{\gamma}_e |\mathbf{h}^{bT}\mathbf{h}^{e*}|^2 \bigg] = 0.
\end{align}


Let $\{\mu_n\}_{n=1}^N$ denote the $N$ roots of \eqref{root_of_eigenvalue}. By the quadratic-root formula, we have 
\begin{align}\label{root}
\mu_{N-1} = 1 + \frac{b - \sqrt{b^2 + 4ac}}{2a}, \quad 
\mu_N = 1 + \frac{b + \sqrt{b^2 + 4ac}}{2a},
\end{align}

According to the Cauchy criterion, it can be easily obtained $c\geq0$. Consequently, $(\mu_{N-1} - 1)(\mu_N - 1) = \frac{-c}{a} \leq 0$, which implies $\mu_N >1$ is the principal eigenvalue of matrix $\mathbf{\Delta}$, i.e., $\mu_{\mathbf{\Delta}}=\mu_N$. The corresponding optimal secrecy rate is given by $R^{sec}=\log_2\left(1 + \frac{b + \sqrt{b^2 + 4ac}}{2a}\right)$. Thus, the final results follow immediately.

\section{The derivative of $f(x_n^p)$ with respect to $x_n^p$}\label{Appendix_B}
To facilitate differentiation, we first present the explicit form of the following terms,
\begin{subequations}
\begin{align}
&a(x_n^p) = 1+\hat{\gamma}_e \|\mathbf{h}^e(x_n^p)\|^2, \\
&b(x_n^p)=\hat{\gamma}_b \|\mathbf{h}^b(x_n^p)\|^2-\hat{\gamma}_e \|\mathbf{h}^e(x_n^p)\|^2\nonumber\\
&\quad~~~~+\hat{\gamma}_b\hat{\gamma}_e \|\mathbf{h}^b(x_n^p)\|^2\|\mathbf{h}^e(x_n^p)\|^2-\hat{\gamma}_b\hat{\gamma}_e \rho_{b,e}(x_n^p), \\
&c(x_n^p)=\hat{\gamma}_b\hat{\gamma}_e \|\mathbf{h}^b(x_n^p)\|^2\|\mathbf{h}^e(x_n^p)\|^2  - \hat{\gamma}_b\hat{\gamma}_e \rho_{b,e}(x_n^p), \\
&\|\mathbf{h}^i(x_n^p)\|^2=\bigg(|h^i_n(x_n^p)|^2+\sum_{n'\neq n}|h^i_{n'}(x_{n'}^p)|^2\bigg),~i\in\{b,e\}, \\
&\rho_{b,e}(x_n^p)=\big|\mathbf{h}^{bT}(x_n^p)\mathbf{h}^{e*}(x_n^p)\big|^2\nonumber\\
&\quad~~~~=\big|h^b_n(x_n^p)h^{e*}_n(x_n^p)+\sum_{n'\neq n}h^b_{n'}(x_{n'}^p)h^{e*}_{n'}(x_{n'}^p)\big|^2.
\end{align}
\end{subequations}
Applying the fundamental rule of fractional differentiation, the derivative of $f(x_n^p)$ with respect to $x_n^p$ is given by:
\begin{align}\label{dfdx}
&\nabla_{x_n^p} f(x_n^p) = \frac{1}{a(x_n^p)^2} \Bigg[ a(x_n^p) \cdot \frac{db(x_n^p)}{dx_n^p} \nonumber\\
& + \frac{a(x_n^p) \left( b(x_n^p) \frac{db(x_n^p)}{dx_n^p} + 2a(x_n^p)  \frac{dc(x_n^p)}{dx_n^p} + 2c(x_n^p) \frac{da(x_n^p)}{dx_n^p} \right)}{\sqrt{b(x_n^p)^2 + 4a(x_n^p)c(x_n^p)}} \\
&\quad - \left( b(x_n^p) + \sqrt{b(x_n^p)^2 + 4a(x_n^p)c(x_n^p)} \right) \cdot \frac{da(x_n^p)}{dx_n^p} \Bigg],\nonumber
\end{align}
where
\begin{align}\label{dadx}
\frac{d a(x_n^p)}{d x_n^p} = -\frac{2 \hat{\gamma}_e \eta (x_n^p - x^e)}{D_n^{e4}},
\end{align}
\begin{align}\label{dbdx}
\begin{aligned}
&\frac{d b(x_n^p)}{d x_n^p} = \frac{2 \hat{\gamma}_e \eta (x_n^p - x^e)}{D_n^{e4}} -\frac{2 \hat{\gamma}_b \eta (x_n^p - x^b)}{D_n^{b4}}- \hat{\gamma}_b \hat{\gamma}_e \frac{d \rho_{b,e}(x_n^p)}{d x_n^p}\\
& - \frac{2 \hat{\gamma}_b \hat{\gamma}_e \eta (x_n^p - x^b) \|\mathbf{h}_e(x_n^p)\|^2}{D_n^{b4}} 
- \frac{2 \hat{\gamma}_b \hat{\gamma}_e \eta (x_n^p - x^e) \|\mathbf{h}_b(x_n^p)\|^2}{D_n^{e4}},
\end{aligned}
\end{align}
\begin{align}\label{dcdx}
\frac{d c(x_n^p)}{d x_n^p}& = 
-\frac{2 \hat{\gamma}_b \hat{\gamma}_e \eta (x_n^p - x^b) \|\mathbf{h}_e(x_n^p)\|^2}{D_n^{b4}}\nonumber\\ 
&- \frac{2 \hat{\gamma}_b \hat{\gamma}_e \eta (x_n^p - x^e) \|\mathbf{h}_b(x_n^p)\|^2}{D_n^{e4}} 
- \hat{\gamma}_b \hat{\gamma}_e \frac{d \rho_{b,e}(x_n^p)}{d x_n^p},
\end{align}
and
\begin{align}
D_n^i = \|&\boldsymbol{\psi}^i - \boldsymbol{\psi}_n^p\| = \sqrt{(x^i - x_n^p)^2 + (y^i - y_n^p)^2 + d^2},\nonumber\\
&~i\in\{b,e\}.
\end{align}
Let $S(x_n^p)=h^b_n(x_n^p)h^{e*}_n(x_n^p)+C$, where $C=\sum_{n'\neq n}h^b_{n'}(x_{n'}^p)h^{e*}_{n'}(x_{n'}^p)$. Then, $\frac{d\rho_{b,e}(x_n^p)}{dx_n^p}=\frac{d\left| S(x_n^p) \right|^2}{dx_n^p}  = 2 \Re \left[ S^*(x_n^p) \frac{d S(x_n^p)}{dx_n^p} \right]$,
which yields that
\begin{align}
\frac{d \rho_{b,e}(x_n)}{d x_n} &= 2 \Re \Bigg[ \left(h^{b*}_n(x_n^p)h^{e}_n(x_n^p)+C^*\right) \nonumber\\
&\left( \frac{d h^b_n(x_n^p)}{d x_n^p} h^{e*}_n(x_n^p) + h^b_n(x_n^p) \frac{d h^{e*}_n(x_n^p)}{d x_n^p} \right) \Bigg].
\end{align}
For simplicity, $h^i_n(x_n^p) = \frac{\sqrt{\eta} \, e^{-j \phi_i}}{D_n^i}, \phi_i = 2 \pi \left( \frac{D_n^i}{\lambda_c} + \frac{\|\boldsymbol{\psi}_{0,n}^p - \boldsymbol{\psi}_n^p\|}{\lambda_p}  \right ), ~ i\in\{b,e\}$.
Therefore,
\begin{align}\label{dhdx}
\begin{aligned}
&\frac{d h^i_n(x_n^p)}{dx_n^p} = \sqrt{\eta} \, e^{-j \phi_i} \Bigg( j 2 \pi \left( \frac{x^i - x_n^p}{\lambda_c D_n^{i2}} - \frac{1}{\lambda_pD_n^i}\right ) - \frac{x_n^p - x^i}{D_n^{i3}}\Bigg ),\\
&\frac{dh^{i*}_n(x_n^p)}{dx_n^p} = \sqrt{\eta} \, e^{j \phi_i} \left(j 2 \pi \left( \frac{x_n^p - x^i}{\lambda_c D_n^{i2}} + \frac{1}{\lambda_pD_n^i}\right ) - \frac{x_n^p - x^i}{D_n^{i3}} \right ).
\end{aligned}
\end{align}
Substituting \eqref{dadx}-\eqref{dhdx} into \eqref{dfdx} yields the derivative of $f(x_n^p)$.

\newpage

%
%
%
%

\vfill

\end{document}